\documentclass[twocolumn]{aastex631}

\usepackage{multirow}
\usepackage{makecell}

\DeclareRobustCommand{\ion}[2]{%
\relax\ifmmode
\ifx\testbx\f@series
{\mathbf{#1\,\mathsc{#2}}}\else
{\mathrm{#1\,\mathsc{#2}}}\fi
\else\textup{#1\,{\mdseries\textsc{#2}}}%
\fi}
\NewCommandCopy\oldtextbf\textbf

\newcommand\nh{\ifmmode{n_{\tiny \mbox{H}}}\else{$n_{\tiny \mbox{H}}$}\fi}
\newcommand\Nh{\ifmmode{N_{\tiny \mbox{H}}}\else{$N_{\tiny \mbox{H}}$}\fi}
\newcommand\ngr{\ifmmode{n_{\tiny \mbox{gr}}}\else{$n_{\tiny \mbox{gr}}$}\fi}
\newcommand\jvsp{\ifmmode{j_{\tiny \nu,sp}}\else{$j_{\tiny \nu, sp}$}\fi}
\newcommand\jvff{\ifmmode{j_{\tiny \nu,ff}}\else{$j_{\tiny \nu,ff}$}\fi}
\newcommand\jvbb{\ifmmode{j_{\tiny \nu,bb}}\else{$j_{\tiny \nu,bd}$}\fi}
\newcommand\amax{\if{a_{\tiny \mbox{max}}}\else{$a_{\tiny \mbox{max}}$}\fi}
\newcommand\amin{\if{a_{\tiny \mbox{min}}}\else{$a_{\tiny \mbox{min}}$}\fi}
\newcommand\cmvol{\ifmmode{\mbox{cm}^{-3}}\else{$\mbox{cm}^{-3}$}\fi}

\submitjournal{ApJ}

\shorttitle{radio--FIR Galaxy SED Reloaded}
\shortauthors{Yoon et al.}

\begin{document}

\title{Prospects for Observing Galaxy Spectral Energy Distribution from the Radio to the far-Infrared in the Era of Next-Generation Radio Telescopes}

\correspondingauthor{Ilsang Yoon}
\email{iyoon@nrao.edu}

\author[0000-0001-9163-0064]{Ilsang Yoon}
\affiliation{National Radio Astronomy Observatory, 520 Edgemont Road, Charlottesville, VA 22903, USA}
\affiliation{Department of Astronomy, University of Virginia, P.O. Box 3818, Charlottesville, VA 22903, USA}

\author{Jonathan Letai}
\affiliation{Department of Physics, Northeastern University, 100 Forsyth St, Boston, MA 02115, USA}
\affiliation{Department of Physics, Cornell University, 109 Clark Hall, Ithaca, NY 14853, USA}

\author[0000-0003-1436-7658]{Hansung B.\ Gim} 
\affiliation{Eureka Scientific, 2452 Delmer Street, Suite 100, Oakland, CA 94602, USA}
%\email{hansung.b.gim@gmail.com}

\author[0000-0002-0786-7307]{Eric F. Jiménez-Andrade}
\affiliation{Instituto de Radioastronomía y Astrofísica, Universidad Nacional Autónoma de México, Antigua Carretera a Pátzcuaro \# 8701,\\ Ex-Hda. San José de la Huerta, Morelia, Michoacán, México C.P. 58089}

\author[0000-0003-1187-4240]{Intae Jung}
\affiliation{Space Telescope Science Institute, 3700 San Martin Drive, Baltimore, MD 21218, United States}
%\email{ijung@stsci.edu}

\author{Caitlin Casey}
\affiliation{Department of Physics, University of California, Santa Barbara, 93106, USA}

\author[0000-0001-7089-7325]{Eric J. Murphy}
\affiliation{National Radio Astronomy Observatory, 520 Edgemont Road, Charlottesville, VA 22903, USA}
\affiliation{Department of Astronomy, University of Virginia, P.O. Box 3818, Charlottesville, VA 22903, USA}

\author[0000-0001-7095-7543]{Min S. Yun}
\affiliation{Department of Astronomy, University of Massachusetts, Amherst, MA 01003, USA}

\begin{abstract}
The superb sensitivity and angular resolution of the next-generation radio telescopes with combined frequency coverage of approximately over three orders of magnitude (100 MHz--100 GHz) will sample the radio and far-infrared (FIR) spectral energy distribution (SED) of galaxies and revolutionize the galaxy formation study at the epoch of re-ionization and beyond. We present a prospect of observing the radio--FIR continuum SEDs of galaxies in the redshift of up to $z\approx 20$ based on an ensemble of the simulated `energy balanced' panchromatic SED (from UV to FIR) extended to the radio. For `realistic' populations of UV star-forming galaxies and dusty star-forming galaxies, we simulate their SEDs by accounting for the CMB effect and the radio--IR correlation. The flux density evolution of the UV-bright star-forming galaxies and the dusty star-forming galaxies at the selected observing frequencies covered by the current (ALMA) and next generation (SKA and ngVLA) radio-millimeter telescopes, suggest that massive galaxies (M$_* \gtrsim 10^{10}$M$_{\odot}$) are detectable at any redshift ($0<z<20$) in high frequency ($\nu>90$GHz). In particular, when operating, the ngVLA high-frequency ($\approx 100$ GHz) band is capable of detecting galaxies with M$_* \gtrsim 10^{9}$M$_{\odot}$ almost independently from redshift and the SKA low-frequency observing window ($\lesssim1$ GHz) has sufficient sensitivity to detect M$_* \gtrsim 10^{10}$M$_{\odot}$ dusty star-forming galaxies up to the epoch of reionization ($z=5\sim7$). We also show that the brightness of anomalous microwave emission (AME) in the galaxy SED is insignificant if the galaxies are beyond the local Universe (e.g., $z\gtrsim 0.1$).
\end{abstract}


\section{Introduction}\label{sec:intro}
The James Webb Space Telescope (JWST) has opened up a new era with the discovery of many UV-bright (M$_{\mbox{\tiny UV}}\lesssim-21$ mag) high redshift ($z\gtrsim10$) galaxies that are either confirmed spectroscopically \citep{arrabal_etal_2023,borsani_etal_2023,bunker_etal_2023,harikane_etal_2023} or inferred photometrically \citep{adams_etal_2023,atek_etal_2023,casey_etal_2023,castellano_etal_2022,donnan_etal_2023,finkelstein_etal_2022,furtak_etal_2023,naidu_etal_2022,rodighiero_etal_2023,santini_etal_2023,topping_etal_2022,whitler_etal_2023,yan_etal_2023}. The continuously growing sample of high-$z$ galaxies will provide important constraints on the galaxy formation models. However, JWST observation of the rest-frame UV emission from those high-$z$ galaxies probes only the young stellar population and misses the interstellar medium, which are other major constituents of galaxies and can be probed by radio--far-Infrared observation.

Thanks to the negative $K$-correction, observations of the far-infrared (FIR) and millimeter continuum spectral energy distribution (SED) have been used to detect the so-called sub-\textit{mm} galaxies at high redshift \citep[For review, see][]{blain_etal_2002}. However, these studies focused on the extreme star-forming galaxy population, which is bright in the FIR--\textit{mm} wavelength, and normal galaxies with a moderate star formation rate (SFR) received little attention.

The recent JWST observations revolutionize the landscape of high-$z$ galaxy observation: the rest-frame optical spectrum is easily obtained for galaxies with $z\sim$ 6--9 and even higher-redshift ($z>10$) galaxies have been discovered, although the majority of them are still based on photometric redshift. Many of these high-$z$ galaxies are normal star-forming galaxies with SFR$\sim$1--10 $M_{\odot} \mbox{yr}^{-1}$ and are considered to be responsible for the cosmic re-ionization \citep[e.g.,][]{robertson_etal_2015,robertson_2022}. It is therefore important to study the properties of gas and dust in these ``normal’’ star-forming galaxies at high redshift.

The high sensitivities and resolutions of the radio--millimeter facilities like ALMA, ngVLA and SKA will play an important role in the observation of radio and FIR continuum emission from such galaxies through a wide range of redshifts. Thanks to the shift of the rest-frame FIR SED peak into their observing frequencies, these facilities are expected to observe the Rayleigh-Jeans tail of the modified black body emission from such galaxies up to the epoch of re-ionization and beyond, which complements the observations from JWST and other coming optical and near-IR observing facilities. For example, recent ALMA observations detect thermal dust continuum emission from star forming main sequence galaxies in $z=6\sim7$ \citep[e.g.,][]{inami_etal_2022,leeuwen_etal_2025}. It is therefore necessary and timely to perform an in-depth examination of the feasibility of the radio--FIR continuum observations with current and future radio telescopes \citep[for example, see][for high-$z$ AGN detection in radio observation]{latif_etal_2025a,latif_etal_2025b}. This paper seeks to investigate the redshift variation of the observed flux density for two major galaxy populations: UV-bright star-forming galaxies and IR-bright, dusty star-forming galaxies (DSFGs), at several key observing frequencies covered by the current (ALMA) and the next-generation (SKA and ngVLA) telescopes. For this investigation, we explicitly consider the effect of the cosmic microwave background (CMB) on the SED shape, which becomes more and more significant with increasing redshift.

The motivation for this investigation should be clear: (1) detecting the thermal dust continuum emission can provide constraints on early galaxies’ physical properties such as the dust temperature and mass, as well as improve the photometric redshift of the high-$z$ candidates which will routinely be produced by JWST; (2) the advent of the JWST era and the capabilities achievable by current and next-generation radio--millimeter telescopes make it imperative that we understand the potential of these continuum radio--FIR observations in order to maximize their scientific value. 

The outline of this paper is as follows: in Section~\ref{sec:overview}, we provide a brief overview of the major radiation processes that dominate the galaxy SED in radio--FIR; Section~\ref{sec:sim} describes the methods used to carry out the investigation; Section~\ref{sec:galpop} introduces the population of galaxies we are simulating; the results are given in Section~\ref{sec:result}; and, lastly, Section~\ref{sec:summary} summarizes our findings. Although the calculations presented in this paper do not require the explicit use of a particular cosmology, it is noted that the data used were all created assuming a $\Lambda$CDM cosmology with $H_0=70\text{ km/s/Mpc}$, $\Omega_M=0.3$, and $\Omega_\Lambda=0.7$.

\section{Radiative processes in the radio--FIR SED} \label{sec:overview}
\begin{figure}
\includegraphics[width=0.5\textwidth]{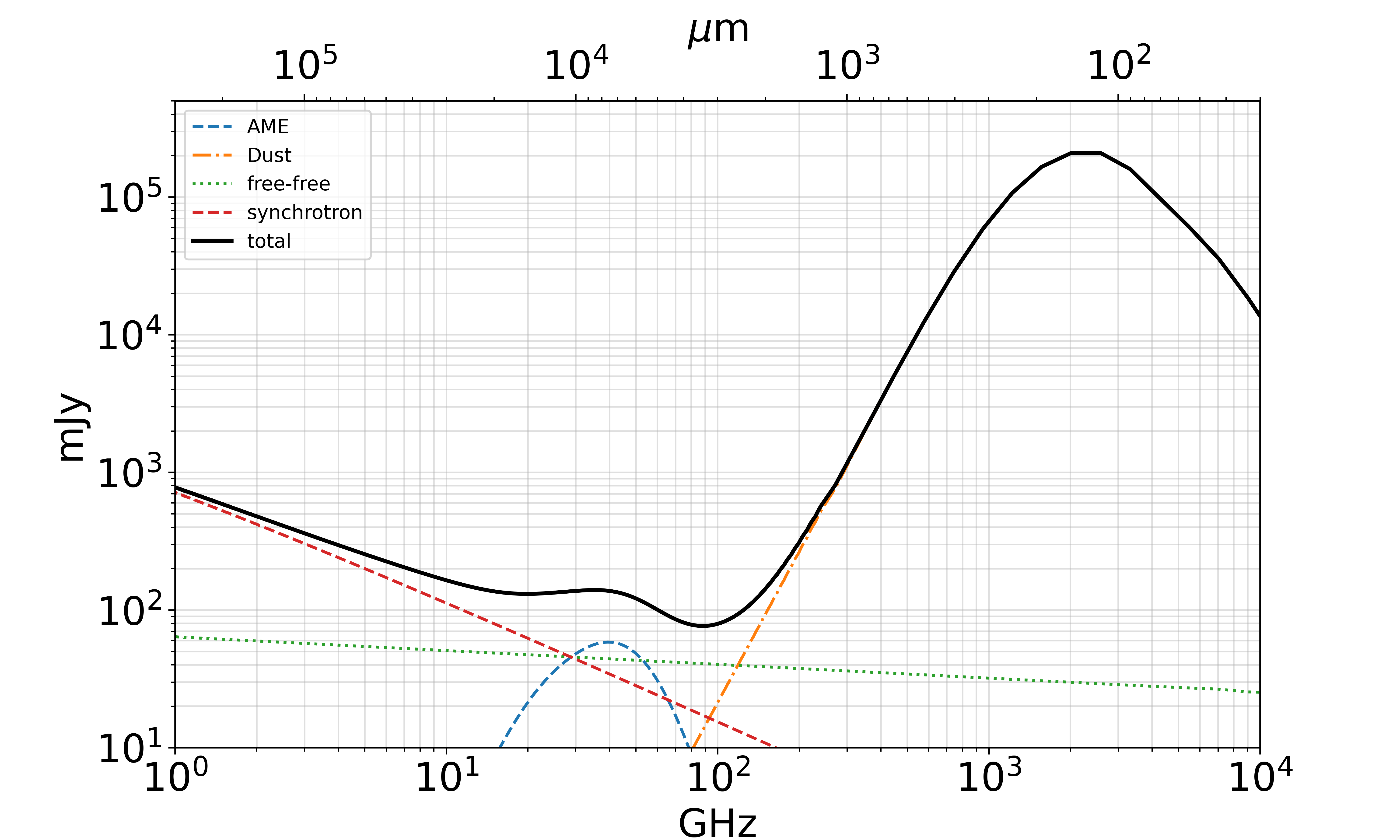}
\caption{Model galaxy spectral energy distribution illustrating four major radiation processes: synchrotron emission, free-free emission, anomalous microwave emission, and thermal dust emission }\label{fig:sed_illustration}
\end{figure}

The focus of this study is to investigate how the galaxy's radio--FIR continuum SED\footnote{Hereafter, the word `SED' means continuum SED.} is determined by the relative contributions from different radiation processes and how they vary as a function of redshift as the CMB effect becomes more significant with increasing redshift. For our study, we implicitly assume that there is no contribution from strong FIR emission lines (e.g., CO and [CII]) to the continuum flux density at the chosen frequencies for our investigation. Indeed, the radio-FIR continuum observation setup has a wide bandwidth and there is a suitable range of frequency for continuum flux measurement while avoiding emission lines. In Figure~\ref{fig:sed_illustration}, we present a model SED of galaxy including four constituents: synchrotron emission, thermal free-free emission, anomalous microwave emission, and thermal dust emission, to illustrate how the galaxy radio-FIR SED are shaped by these radiation processes. Radio continuum emission is dominated by synchrotron emission in low-frequency ($\nu \lesssim 1$GHz) and by thermal dust emission in high-frequency ($\nu \gtrsim 100$ GHz) while the radio emission in the intermediate-frequency (1--100 GHz) is set by a mixture of the four radiation processes. 

The model spectrum in Figure~\ref{fig:sed_illustration} serves as an illustration of a galaxy SED and does not include more sophisticated physical processes such as a spectral flattening in low frequency ($\nu<1$ GHz) observed from a large sample of star-forming galaxies, which increases with V-band optical depth and is probably due to the energy loss of cosmic ray electrons and free-free absorption \citep{an_etal_2024}. The detailed investigation of the impact of spectral flattening on the radio-FIR SED is beyond the scope of this paper and we do not include the physics of the low-frequency spectral flattening in our simulation because a physical model or an empirical prescription is not well developed. However, we note that the spectral flattening in the rest-frame $<1$ GHz may have an impact on our model SED only in the frequency range 0.11--0.77 GHz for nearby galaxies ($z<1$) while the impact is negligible for the majority of high-$z$ galaxies ($z>3$). Furthermore, UV star-forming galaxies are not detectable in low-frequency radio bands (0.11--0.77 GHz) due to the low sensitivities even before we consider the spectral flattening, as discussed in Section~\ref{subsubsec:jaguargal}. Therefore the spectral flattening in low frequency radio bands does not have a significant impact in our study. Before we present the result of our investigation (Section~\ref{sec:result}), in this section, we provide a brief overview of the radiation processes relevant to the radio--FIR SED of galaxies. 

\subsection{Synchrotron emission}\label{subsec:synchphysics}
Synchrotron emission originates from cosmic ray (CR) electrons accelerated in a magnetic field and dominates the radio SED for most normal galaxies at frequencies below $\nu\sim10$GHz, with a typical spectral power-law index of $-0.8$ \citep[$S\sim\nu^{-0.8}$,][]{condon_1992}. If anomalous microwave emission is not detected, synchrotron emission determines the spectral shape of radio emission at $\nu \lesssim 10$ GHz although the radio SED at low frequency ($\nu \lesssim1$ GHz) may be flattened due to CR electron energy loss or free-free absorption \citep{an_etal_2024} which we do not consider in our investigation as explained above. The source of CR injection can be a supernova explosion in star-forming regions or active galactic nuclei (AGN). As CR electrons propagate through the ISM of galaxies, they cool via several processes: synchrotron radiation, inverse Compton (IC) scattering with CMB as well as ionization, bremsstrahlung, and adiabatic expansion \citep{murphy_2009}. 

A recent study by \cite{yoon_2024} compares the energy loss rate of each cooling process for CR electron ($P~[\mbox{erg s$^{-1}$}]$) and estimates the fraction of synchrotron cooling relative to the total energy loss rate 
\begin{equation}
f_{\mbox{\tiny synch}} = \frac{P_{\mbox{\tiny synch}}}{P_{\mbox{\tiny CR}}}\label{eq:sync_frac}
\end{equation}
where $P_{\mbox{\tiny CR}}$ is the sum of energy loss rate of high energy CR electron due to synchrotron emission ($P_{\mbox{\tiny synch}}$), IC scattering with CMB ($P_{\mbox{\tiny IC}}$), bremsstrahlung emission ($P_{\mbox{\tiny brem}}$), ionization ($P_{\mbox{\tiny ion}}$), and adiabatic expansion ($P_{\mbox{\tiny ad}}$): $P_{\mbox{\tiny CR}}
=  P_{\mbox{\tiny synch}} + P_{\mbox{\tiny IC}} + P_{\mbox{\tiny brem}} + P_{\mbox{\tiny ion}} + P_{\mbox{\tiny ad}} \quad[\mbox{erg s$^{-1}$}]$. The synchrotron energy loss rate depends on the magnetic field energy density $U_{\mbox{\tiny B}}$ \citep{rybicki_and_lightman_1979}
\begin{equation}
P_{\mbox{\tiny synch}}=\frac{4}{3}\sigma_{\mbox{\tiny T}}c\gamma^2\beta^2 U_{\mbox{\tiny B}}\label{eq:psynch}
\end{equation} where $\sigma_{\mbox{\tiny T}}$, $c$, $\gamma$, and $\beta$ are respectively, Thompson cross section of electron, speed of light, Lorentz factor, and the ratio between the electron speed and speed of light. For a fixed galaxy property, the synchrotron cooling fraction (Equation~\ref{eq:sync_frac}) for the galaxy is expected to be smaller with increasing redshift due to the increasing efficiency of CMB IC cooling, resulting in `synchrotron dimming' for the high-$z$ galaxies as suggested in the previous studies \citep[e.g.,][]{lacki_etal_2010b,murphy_2009}. The synchrotron dimming impacts the radio SED shape and lowers the radio flux density as we move a model galaxy from low- to high-$z$ Universe. Observationally, this implies that the radio--IR correlation well established in local Universe \citep{helou_etal_1985,yun_etal_2001} becomes steeper for the same galaxies if they are located in higher redshift. However, the radio--IR correlation for high-$z$ galaxies does not show a systematic change along the redshift \citep{murphy_2009,ivison_etal_2010,delvecchio_etal_2021}. 

To explain the observed radio--IR correlation as a function of redshift, \cite{yoon_2024} proposes a simple model by considering the amplification of magnetic field \citep[e.g., ][]{schleicher_and_beck_2013} in the turbulent ISM in the high-redshift Universe. It is shown that small-scale turbulent dynamo can amplify the seed magnetic field \citep{kazantsev_1968,kulsrud_and_anderson_1992,brandenburg_and_subramanian_2005} and increase the turbulent magnetic field strength through the turbulent energy injection from supernova explosions \citep[e.g.,][]{gent_etal_2023,rieder_etal_2017,pakmor_etal_2024,schober_etal_2013}. If the magnetic field is increased in turbulent ISM, the turbulent magnetic field in high-$z$ galaxies due to a strong feedback process is indeed expected to be higher than the strength of magnetic field in the local Universe. 

The fraction of synchrotron cooling relative to other CR electron cooling processes (measured by $f_{\mbox{\tiny synch}}$) deviates from the value established in the local universe by a factor of $f_s$\citep{yoon_2024} 
\begin{eqnarray}\label{eq:fscale}
    f_s & = & \frac{f_{\mbox{\tiny synch}}(z)}{f_{\mbox{\tiny synch}}(0)}\\\nonumber
    & = & \frac{P_{\mbox{\tiny synch}}(0) + P_{\mbox{\tiny IC}}(0) + P_{\mbox{\tiny brem}}(0) + P_{\mbox{\tiny ion}}(0) + P_{\mbox{\tiny ad}}(0)}{P_{\mbox{\tiny synch}}(z) + P_{\mbox{\tiny IC}}(z) + P_{\mbox{\tiny brem}}(z) + P_{\mbox{\tiny ion}}(z) + P_{\mbox{\tiny ad}}(z)}\\\nonumber
    & & \times\frac{P_{\mbox{\tiny synch}}(z)}{P_{\mbox{\tiny synch}}(0)}.
\end{eqnarray} where each $P$ is energy loss rate (erg/s) associated with different cooling processes: synchrotron, CMB inverse Compton, bremsstrahlung, ionization, and adiabatic expansion. The detailed calculation of the energy loss rate for each cooling process and the $f_s$ value is described in \cite{yoon_2024}. This factor, $f_s$ is multiplied to the synchrotron emission component for a given redshift when we create the radio SED (see Section~\ref{subsec:radsed}). 

In Section~\ref{subsec:radsed}, we discuss two simulations of the radio SED, one including only the synchrotron dimming and the other including the synchrotron dimming with the magnetic field amplification. 

\subsection{Thermal free-free emission}\label{subsec:ffphysics}
The free-free emission in galaxy SED originates by the bremsstrahlung process from thermal motions of electrons in \ion{H}{ii} region and dominates the radio SED at higher frequencies ($\nu\gtrsim30$ GHz) with a typical spectral power-law index of $-0.1$ \citep[$S\sim\nu^{-0.1}$][]{condon_1992}. Attenuation by free-free absorption is generally negligible in the ISM for $\nu\gtrsim10$GHz and the free-free emission can be optically thick for $\nu\lesssim1$ GHz \citep{draine_2011}. Therefore at high-redshift, the rest-frame frequencies SKA observes will not be much affected by free-free absorption \citep{lacki_etal_2010b}. 

The contribution of free-free emission to the radio SED (i.e., thermal fraction) varies depending on the observing frequency, for example, $\approx 10$\% at 1.4GHz \citep{condon_1992,condon_and_yin_1990,niklas_etal_1997} and $\approx 90$\% at 33GHz \citep{murphy_etal_2011,murphy_etal_2018a}. The thermal fraction should be derived from the SED decomposition. Although the flat spectral shape of free-free emission should be distinguishable from the steeper non-thermal spectrum from synchrotron emission, it is difficult to isolate because normal galaxies are not bright enough to be detected at frequencies much higher than 10GHz \citep{condon_1992}.  As we discuss in Section~\ref{sec:sim}, we will simply assume 10\% thermal fraction at 1.4GHz for the use of radio--IR correlation. 

\subsection{Anomalous Microwave Emission}\label{subsec:amephysics}
Anomalous Microwave Emission (AME) explained by dipole radiation from spinning nanoparticle \citep{draine_and_lazarian_1998} may appear as an excess emission in 10--30GHz frequency and the peak frequency can be as high as 100GHz \citep[][]{spdust2009}. The origin and career of the AME and the physical conditions of the ISM for AME are still not clearly understood \citep[][for review]{dickinson_etal_2018}. 

In particular, AME from the extragalactic star-forming region is very poorly understood. While the current estimate of the fraction of AME to the total emission at frequencies near $\approx 30$ GHz in our Galaxy is as large as 50\% \citep{dickinson_etal_2018}, in other galaxies, AME does not seem to be strong, as shown by the observations of two galaxies, NGC6946 \citep{murphy_etal_2010,scaife_etal_2010,hensley_etal_2015} and NGC 4725B \citep{murphy_etal_2018b}. A recent analysis of the AME from several nearby galaxies suggests that the galaxy integrated AME is weak \citep{correia_etal_2026} although a significant AME is reported from the galaxy integrated SED for M31 \citep{battistelli_etal_2019}. It is not clear whether this discrepancy is due to a different ISM environment associated with the AME region in our Galaxy and in other galaxies, or just a bias in the AME estimate in our Galaxy based on the observation of solar neighbor or a combination of both \citep{dickinson_etal_2018}.

Recently, \cite{yoon_2022} investigated the grain disruption process by centrifugal force due to the increased spin-angular momentum of small ($<10^{-7}$cm in size) grains impacted by magnetic shocks \citep{hoang_etal_2019} and compared its strength relative to the other competing emission mechanisms in the star-forming region: thermal free-free and dust continuum emission. It was demonstrated that, if the ISM in a magnetic field is impacted by a C-shock developed by a supernova explosion in the early phase of massive star-formation ($\lesssim 10$ Myr), AME can be significantly or almost entirely suppressed relative to free-free and thermal dust continuum emission if the centrifugal force of grains due to the increased spin angular momentum caused by the stochastic mechanical torque from the C-shock is stronger than the grain tensile strength. This study may shed light on explaining the rare observations of AME from extragalactic star-forming regions preferentially observed from massive star clusters and suggest a scenario of ``the rise and fall of AME'' in accordance with the temporal evolution of star-forming regions \citep{yoon_2022}.

As we will show later in this study, the next-generation radio telescopes: SKA and ngVLA will have sufficient resolution and sensitivity to detect the emission from many individual extragalactic star-forming regions in galaxies in the nearby and distant Universe. In a massive star-forming region with a strong radiation field, the centrifugal disruption of large grains due to the radiative torque \citep{hoang_etal_rat_2019} increases a relative fraction of small grains which increases the brightness of AME as long as the star-forming region is still young and not yet impacted by C-shock. More interestingly, the CMB radiation field in high-$z$ Universe can be a significant or even a dominant contributor to the radiation field that excites the rotation of grains. Therefore the AME can be a new interesting aspect of the radio SED that may be more frequently detected in the era of SKA and ngVLA. Since the AME is a phenomenon depending on local ISM conditions and therefore is not appropriate to incorporate into the global galaxy SED model, in Section~\ref{sec:sim}, we will present the AME SED and its variation with redshift, independently from other SED components. 

\subsection{Thermal dust emission}\label{subsec:thermalphysics}
The FIR SED of a galaxy is dominated by the thermal blackbody emission from dust modified by the dust opacity with a power-law index $\beta$ ($1<\beta<2$). The steep spectral shape ($S\sim\nu^{2+\beta}$) of the SED in the Rayleigh-Jeans tail has been recognized as a powerful observing tool to detect distant galaxies via strong `negative \textit{K}-correction' \citep{blain_and_longair_1993,blain_etal_2002}. However, with increasing redshift, CMB impacts the observation of thermal dust emission in two ways: CMB affects the FIR SED by enhancing the thermal blackbody component of the FIR SED (i.e., heating) but also by reducing the contrast of the detection (i.e., strong background emission). 

Although the CMB effect on the FIR SED has been investigated with the suggested prescription for correcting the effect \citep{dacunha_etal_2013}, it has not been incorporated in the creation of a full panchromatic SED model. In this study, we follow the procedure in the appendix of \cite{yoon_etal_2023} when simulating the galaxy SED, accounting for the effect of the CMB on the FIR model SED component and predicting an `observed' FIR SED at any given redshift.

As mentioned in Section~\ref{subsec:modeling}, thermal dust emission in FIR is a reprocessed UV stellar light absorbed by dust. Therefore, the FIR SED also depends on the dust properties (e.g., mass, temperature grain size, opacity). Compared to the local galaxies, the properties of dust in high-$z$ galaxies are less well understood although several studies investigated the evolution of dust temperature \citep[e.g.,][]{mitsuhashi_etal_2024a,sommovigo_etal_2022} and mass \citep{mauerhofer_and_dayal_2023} as a function of redshift, and studied the metallicity effect, dust composition and production \citep{boquien_etal_2022,choban_etal_2025,markov_etal_2023,witstok_etal_2023} for high-$z$ galaxies. In particular, the dust properties are not known for very high redshift ($z\gtrsim 8$) galaxies: non-detection of thermal dust emission from galaxies at such high redshift suggests strong stellar feedback blowing the dust out of galaxies \citep{ferrara_etal_2025a,ferrara_etal_2025b} or inefficient dust production \citep{behrens_etal_2018,heintz_etal_2025,toyouchi_etal_2025}. Given that these high-$z$ galaxies are likely to be in an early stage of star-formation, nascent starburst galaxies observed in the local Universe might provide good insight: for example, \cite{roussel_etal_2003} report unusually hot dust with IR-excess in the radio--IR correlation from three nascent local starburst galaxies. If high-$z$ galaxies share similar properties with these local nascent starburst galaxies --- hot dust and IR-excess --- the FIR SED is affected by dust which is hotter than that considered in this study and the radio synchrotron flux density might be fainter for a fixed IR luminosity than the expected value inferred from the conventional radio--IR correlation parameter $q$ we adopt in this study (Section~\ref{subsec:radsed}).

With these caveats on the dust properties of high-$z$ galaxies, we will assume that a widely used prescription of the FIR emission reprocessed by dust via `energy balance' \citep{dacunha_etal_2008} using a representative extinction curve also holds for high-$z$ galaxies although more realistic simulation of FIR dust emission and radio synchrotron emission requires a better understanding of the dust properties in high-$z$ universe. 

\section{The simulation of galaxy spectral energy distribution from the optical/NIR to the FIR/radio} \label{sec:sim}
\subsection{Panchromatic SED model} \label{subsec:modeling}
Simulating a panchromatic galaxy SED from the optical/NIR to the FIR/radio regimes is useful because it provides the physical insight into the relationship between the stars and dust in galaxies and the rate of star formation in both `obscured' and `unobscured' mode. In this work, we simulate the SEDs of galaxies with `realistic' properties motivated by the recent simulations for the JWST Advanced Deep Extragalactic Survey \citep{Williams_2018} and the multiwavelength observations of the high-redshift dusty star-forming galaxies \citep{Liang_2019}. Properties of the stellar population and star formation from these galaxy samples are used as input parameters for the simulation (see Section \ref{sec:galpop}). 

The resulting FIR SED is generated by reprocessing the UV stellar light absorbed by dust into the FIR wavelength, based on the energy balance assumption. Although there is a question of whether the energy balance assumption holds, for example, if UV and FIR emission are not co-spatial, recent quantitative analysis supports the validity of the energy balance assumption for high-redshift galaxies with UV-FIR offsets \citep{haskell_etal_2023}. The radio SED is determined by the typical spectral shape for synchrotron and free-free emission and the flux density based on the IR luminosity using the radio--IR correlation (Section~\ref{subsec:radsed}).  

For simulating galaxy SEDs from UV to radio wavelength, we choose \texttt{CIGALE} \citep{Boquien_2019} amongst several available popular SED modeling tools because it has a parameterized FIR SED model from \cite{Casey_2012} that is useful for this study in terms of controlling the dust temperature when simulating galaxy SED.   

\subsection[CIGALE]{\normalfont{\texttt{CIGALE}}} \label{subsubsec:CIGALE}
Code Investigating GALaxy Emission, \texttt{CIGALE} \citep{Boquien_2019}, utilizes a set of modules to sequentially build up an SED by calculating and adding in different contributions one at a time. The modules in \texttt{CIGALE} used for this study are listed below, with a brief description \citep[see][for a full description]{Boquien_2019}: 
\begin{enumerate}
    \item {\tt{sfhdelayed}} --- Creates a star formation history for each galaxy using a parametric model of the form $SFH(t)=\frac{t}{\tau^2}e^{-\frac{t}{\tau}}\text{ for } 0\leq t\leq t_{*}$, where $t_{*}$ is the age of the onset of star formation.
    \item {\tt{bc03}} --- Computes the stellar spectrum from the SFH and the SSPs described in \citet{BC_2003}. For the initial mass function (IMF), we use a Chabrier IMF \citep{Chabrier_2003}. 
    \item {\tt{dustatt\_modified\_CF00}} --- Calculates the dust attenuation, and thence the luminosity absorbed by the dust, via the \citet{CF_2000} model.
    \item {\tt{casey2012}} --- Computes the dust emission via the \citet{Casey_2012} model while following the principle of energy balance: the energy absorbed by the dust is equal to the energy emitted by the dust. This model is comprised of two additive components: a single-temperature modified blackbody and a power-law function. PAH emission is not accounted for, but should have a negligible effect on the FIR and radio SED. (NOTE: This module is modified for use in this investigation, as described in Section \ref{subsec:firsed}.)
    \item {\tt{radio}} --- Calculates the contribution of synchrotron radio emission by using the radio--IR correlation (described in \citet{helou_etal_1985}) with the IR emission computed above. (NOTE: This module is modified for use in this investigation, as described in Section \ref{subsec:radsed}.)
    \item {\tt{redshifting}} --- Redshifts the spectrum and accounts for the absorption of radiation by the IGM. (NOTE: This module is modified for use in this investigation, as described in Section \ref{subsec:firsed}.)
\end{enumerate}

Lastly, the default normalization in which {\tt{CIGALE}} sets $\int SFH(t)dt=1$ is also used. Thus, the output SED represents a galaxy with stellar mass $M_*$ approximately equal to 1. $M_*$ is not exactly 1, however, because stars return mass to the ISM \citep{Boquien_2019}. The actual stellar mass is calculated in the {\tt{bc03}} module (usually $M_*\approx 0.8M_{\odot}$) and stored in the {\tt{stellar.m\_star}} parameter. It is necessary to account for this correction when scaling the SED to achieve a desired galaxy stellar mass.

\subsection{Radio SED}\label{subsec:radsed}
The radio model SED in \texttt{CIGALE} uses the radio--IR correlation \citep{helou_etal_1985}
\begin{equation}\label{eq:qir}
q\equiv\mbox{log}\left(\frac{F_{\mbox{\tiny IR}}(8-1000\mu\mbox{m})}{3.75\times10^{12}\mbox{W m}^{-2} \mbox{Hz}^{-1}}\right) - \mbox{log}\left(\frac{S_{\mbox{\tiny 1.4GHz}}}{\mbox{W m}^{-2}\mbox{Hz}^{-1}}\right)
\end{equation}
to set the radio flux density at 1.4GHz using the total IR luminosity (integrated over 8--1000 $\mu$m) derived from the energy balance and the assumed 10\% thermal fraction at 1.4GHz, resulting in $q=2.64$ \citep{bell_2003}. We modified the \texttt{radio} module in \texttt{CIGALE} to divide the radio flux into the synchrotron and free-free emission and use the adopted thermal fraction and the spectral indices for synchrotron ($-0.8$) and free-free ($-0.1$) emission. To simulate the synchrotron dimming with increasing redshift, we multiply the scale factor defined by Equation~\ref{eq:fscale} to the synchrotron SED component. 

\begin{figure}
\includegraphics[width=0.5\textwidth]{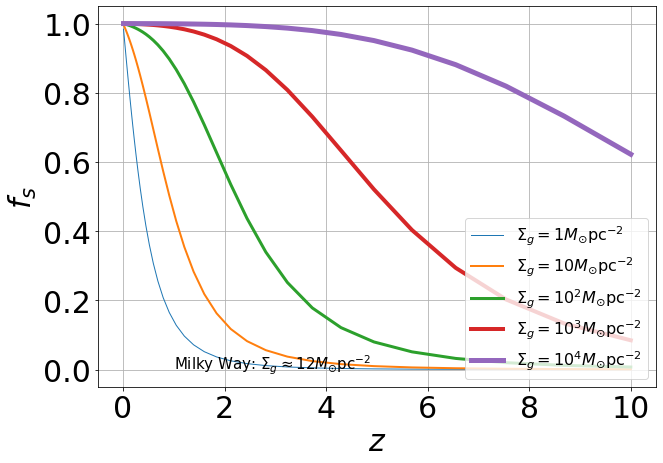}
\caption{Synchrotron dimming factor, $f_s$ in Equation~\ref{eq:fscale} as a function of redshift ($z$) for different gas surface density $\Sigma_g$, without the consideration of turbulent magnetic field enhancement (i.e., $U_{\mbox{\tiny B}}(z)=U_{\mbox{\tiny B}}(0)$)}\label{fig:fscale}
\end{figure}

\cite{yoon_2024} expresses each energy loss rate in Equation~\ref{eq:fscale} as a function of gas surface density\footnote{1 g cm$^{-2}$ = 4800 $M_{\odot}$ pc$^{-2}$ and  $\Sigma_g=12 M_{\odot}$ pc$^{-2}$ is the average value of our Milky Way}($\Sigma_g$) inferred from star formation rate surface density using the Kennicutt-Schmidt relation \citep{ks_relation_2012}. The synchrotron dimming factor ($f_s$ in Equation~\ref{eq:fscale}) depends on the gas surface density and redshift \citep{yoon_2024}. In Figure~\ref{fig:fscale}, we show $f_s$ as a function of redshift for different $\Sigma_g$ without considering the variation of magnetic field energy density as a function of redshift ($U_{\mbox{\tiny B}}(z) = U_{\mbox{\tiny B}}(0)$ in Equation~\ref{eq:psynch}). While the synchrotron emission from a galaxy like our Milky Way with low gas surface density ($\Sigma_g\approx10 M_{\odot}$pc$^{-2}$) becomes weak very quickly with redshift, the suppression of synchrotron emission from star-forming galaxies with a high gas surface density ($\Sigma_g\gtrsim100 M_{\odot}$pc$^{-2}$) is not very significant (i.e., less than 50\% up to $z=2$).    

Although the effect of the CR electron escape is not being considered for simulating radio synchrotron emission, it is unlikely to happen at high enough surface densities \citep[$\Sigma_g \gtrsim 0.1$ g cm$^{-2}$ in ][which is translated to $\Sigma_g \gtrsim 480 M_{\odot}$pc$^{-2}$ or $\Sigma_{\mbox{\tiny SFR}}\gtrsim 0.3$ $M_{\odot}$kpc$^{-2}$yr$^{-1}$]{lacki_etal_2010b}. Even though the escape of CR electron is not explicitly taken into account in the synchrotron dimming, we note that the radio--IR correlation is still largely unchanged because the diffusive escape of CR electrons decreases the radio emission however UV photons also escape without being reprocessed into FIR by absorption \citep[i.e., the ``low-$\Sigma_g$'' conspiracy'' in ][]{lacki_etal_2010b}. 

As discussed in Section~\ref{subsec:synchphysics}, the turbulent magnetic field amplification due to the turbulent energy injection from supernova explosions can enhance the turbulent magnetic field strength as the redshift increases \citep{schleicher_and_beck_2013}, which may compensate the synchrotron dimming in the high-$z$ Universe. If we consider the increased magnetic energy density in high-$z$ galaxies compared to the galaxies in the local Universe by adopting ISM density-dependent magnetic field strength \citep[e.g.,][]{yoon_2024}, the synchrotron emission does not dramatically decrease with redshift when the gas density is high. 

For this study, we simulated an ensemble of galaxy SEDs with two different versions of radio SED: `version 1' with the synchrotron dimming only and `version 2' with the synchrotron dimming and the redshift-dependent magnetic field enhancement, parameterized by gas surface density using $\Sigma_g=10^3 M_{\odot}$ pc$^{-2}$. As we will discuss later, a non-negligible difference in the SED due to the two different synchrotron emission prescriptions (with and without magnetic field enhancement) shows up only in the low-frequency radio SED ($\lesssim$ 6--7 GHz) and does not affect the SED shape at high-frequencies.

Before we present the result in Section~\ref{sec:result}, here we note that there are variations in the radio SED in our simulation due to the choice of gas surface density (i.e., the higher the gas surface density is, the brighter the radio flux density is). The difference in the radio flux density due to the variation of $\Sigma_g$ is noticeable only in low frequency ($\nu \lesssim 10$ GHz).
%and a significant difference (i.e., factor 2 or larger) is only seen in $\nu \lesssim 6$ GHz. 
We emphasize that the predicted radio flux density in low frequency is subject to the assumptions in the creation of the radio SED using IR luminosity. However, for the results presented in this paper, the low-frequency radio SEDs without the effect  of magnetic field enhancement (`version 1') are a conservative lower limit.      

\subsection{FIR SED}\label{subsec:firsed}
We use a parameterized analytic model for FIR emission from \cite{Casey_2012}. Dust emission is modeled with two components: a single-temperature modified blackbody representing the absorbed stellar emission reprocessed by dust in the whole galaxy and a power-law function in the mid-IR that approximates hot-dust emission \citep{Casey_2012}. As mentioned in Section~\ref{subsec:thermalphysics} and also described in the appendix of \cite{yoon_2022}, to consider the increasing CMB effect in the high-redshift Universe, we create a galaxy SED using a graybody FIR spectrum with a dust temperature enhanced by the CMB heating. Based on \cite{dacunha_etal_2013}, the dust temperature, $T_d (z)$ at a given redshift $z$ can be expressed as follows
\begin{equation}\label{eq:at}
     T_d^{4+\beta}(z) = T_d^{4+\beta}(0)-T_{\mbox{\tiny CMB}}^{4+\beta}(0)+T_{\mbox{\tiny CMB}}^{4+\beta}(z)
\end{equation} where $T_{\mbox{\tiny CMB}}(z)$ is the CMB temperature at redshift $z$ and $T_{\mbox{\tiny CMB}}^{4+\beta}(z)=T_{\mbox{\tiny CMB}}^{4+\beta}(0)\times(1+z)^{4+\beta}$ with $T_{\mbox{\tiny CMB}}(0)=2.725$K for the standard $\Lambda$CDM cosmology, and $T_d(0)$ is the dust temperature at $z=0$. If we define a temperature $T_{d,*}\equiv[T_d^{4+\beta}(0)-T_{\mbox{\tiny CMB}}^{4+\beta}(0)]^{1/(4+\beta)}$ as the temperature of the dust at $z=0$ if heated only by star formation, excluding the CMB background heating,
\begin{equation}\label{eq:tdust}
    T_d^{4+\beta}(z)=T_{d,*}^{4+\beta}+T_{\mbox{\tiny CMB}}^{4+\beta}(z)
\end{equation}  where $\beta$ is the dust emissivity with the default value of 1.6 in \texttt{casey2012} in \texttt{CIGALE}. 

In this work, $T_{d,*}$ is considered as a reference temperature of dust heated by star formation with no contribution from CMB heating and we will use a range of $T_{d,*}$ values for simulating FIR SED. For a given choice of $T_{d,*}$ value, we use $T_{d}(z)$ value as an input parameter for \texttt{CIGALE} FIR SED model at a given redshift $z$ by adding CMB contribution $T_{\mbox{\tiny CMB}}(z)$ (Equation~\ref{eq:tdust}). Since the `initial' model SED from \texttt{CIGALE} with the dust temperature $T_d(z)$ does not include the contribution from CMB heating to the IR luminosity, we `boost' the model SED by adding an additional modified black body spectrum with the CMB temperature at the same redshift and apply the correction factor due to the CMB contrast \citep{dacunha_etal_2013}, $1-\{B_\nu\left[T_{\mbox{\tiny CMB}}(z)\right]/B_\nu\left[T_{d}(z)\right]\}$ to the modified black body component of the `boosted' FIR SED as explained in the appendix of \cite{yoon_etal_2023}.

%
%\begin{figure*}
%\centering
%\includegraphics[width=0.32\textwidth]{emiss_cnm.png}
%\includegraphics[width=0.32\textwidth]{emiss_wnm.png}
%\includegraphics[width=0.32\textwidth]{emiss_pdr.png}
%\caption{Emissivity of AME for typical conditions for cold neutral medium (CNM), warm neutral medium (WNM) and photo-dissociated region (PDR). For each ISM condition, the AME emissivity is computed with additional CMB contribution to the ambient radiation field density that becomes more significant with increasing redshift as colored and annotated for different redshift.}\label{fig:ameemiss}
%\end{figure*}
%\begin{figure*}
%\centering
%\includegraphics[width=0.32\textwidth]{ameflux_cnm.png}
%\includegraphics[width=0.32\textwidth]{ameflux_wnm.png}
%\includegraphics[width=0.32\textwidth]{ameflux_pdr.png}
%\caption{Flux density of AME using Equation~\ref{eq:ameflux} for AME %emissivities in Figure~\ref{fig:ameemiss}.}\label{fig:ameflux}
%\end{figure*}

\subsection{AME SED}\label{subsec:ame}
We use \texttt{SpDust} code \citep{spdust2009} to compute the AME emissivity per hydrogen atom, $j_{\nu}/\nh$ (Jy sr$^{-1}$cm$^2$ per H-atom) and compute the observed flux density, $F_\nu$, by assuming a cylindrical geometry for the AME emitting region with the diameter $l$ and the size $\ell$ along the line-of-sight  
\begin{equation}\label{eq:ameflux}
    F_\nu = \frac{\pi}{4}l^2\nh \ell \left(\frac{4\pi j_\nu}{\nh}\right)\bigg/ 4\pi D_{\mbox{\tiny L}}^2~~[\mbox{Jy}]
\end{equation}
where \nh\ is the hydrogen number density (cm$^{-3}$) and $D_{\mbox{\tiny L}}$ is the luminosity distance of the AME source. Using a radio beam size $\theta$, angular diameter distance $D_{\mbox{\tiny A}}$ ($D_{\mbox{\tiny A}}\theta=l$), hydrogen column density \Nh (i.e., \nh $\ell$), $F_\nu$ can be rewritten as 
\begin{equation}\label{eq:ameflux2}
    F_\nu = \frac{1}{206265^2} \left(\frac{\pi}{4}\right) \frac{1}{(1+z)^4} \left(\frac{\theta}{1\arcsec}\right)^2 \Nh \left(\frac{j_\nu}{\nh}\right)~~[\mbox{Jy}]
\end{equation}
where 1 rad$=$206265\arcsec. 

Equation~\ref{eq:ameflux2} implicitly assumes that the size of AME emitting region is the same as the beamsize ($D_{\mbox{\tiny A}}\theta=l$). For a given observing beam, the apparent size of AME emitting region is generally smaller than the observing beamsize for distant galaxies that is not spatially resolved by the observing beam, while the apparent size of AME emitting region for nearby galaxies may be comparable to or larger than the observing beamsize. In principle, the observed AME flux density depends on the size of AME emitting region relative to the observing beamsize, which requires a consideration of the angular resolution of radio telescope to predict the observed AME flux density (Yoon et al. 2026, SKA II Book, in press). For our investigation of the prospect of radio continuum observation from distant galaxies, we assume that the measurements are spatially integrated and therefore the observing beam is large enough to fully contain the AME region ($D_{\mbox{\tiny A}}\theta \lesssim l$). In this study, we will fix the size of AME emitting region ($l=100$ pc).      

The $j_\nu/\nh$ is computed from \texttt{SpDust} with the parameters of the ISM condition. One of the key parameters that changes the strength of AME emissivity is the energy density of ambient radiation field $\chi$ in the unit of the local interstellar radiation field energy density $u_{\mbox{\tiny ISRF}}$. In high-$z$ Universe, the ambient radiation field intensity increases due to the increasing CMB energy density. Therefore the new ambient radiation field energy density $\chi'$ at redshift $z$ is related to the \texttt{SpDust} input parameter for the radiation field energy density $\chi$ at $z=0$ as follows.
\begin{equation}
    \chi' u_{\mbox{\tiny ISRF}} = \chi u_{\mbox{\tiny ISRF}} + U_{\mbox{\tiny CMB}}(z)
\end{equation}
Therefore,
\begin{equation}\label{eq:ambrad}
    \chi' = \chi + \frac{U_{\mbox{\tiny CMB}}(0)}{u_{\mbox{\tiny ISRF}}}(1+z)^4
\end{equation} which implies that the AME emissivity from star-forming region in galaxies will increase at high-$z$ Universe. By adopting $U_{\mbox{\tiny CMB}}(0)=4.2\times10^{-13}$ erg cm$^{-3}$ and $u_{\mbox{\tiny ISRF}}=1.05\times10^{-12}$ erg cm$^{-3}$ \citep{draine_2011}, we will use $\chi'$ for \texttt{SpDust} parameter when we compute the AME emissivity at redshift $z$.

The other required parameters for \texttt{SpDust} are the dust size distribution and the physical conditions of the ISM including the radiation field intensity parameter $\chi$. We use the dust size distribution from \citep{weingartner_and_draine_2001} which is the default option in \texttt{SpDust} and the typical ISM condition for cold neutral medium (CNM), warm neutral medium (WNM) and photo-dissociated region (PDR) in \cite{draine_2011}. Note that the contribution from CMB to $\chi'$ value increases with $(1+z)^4$.

\begin{figure}
\includegraphics[width=0.5\textwidth]{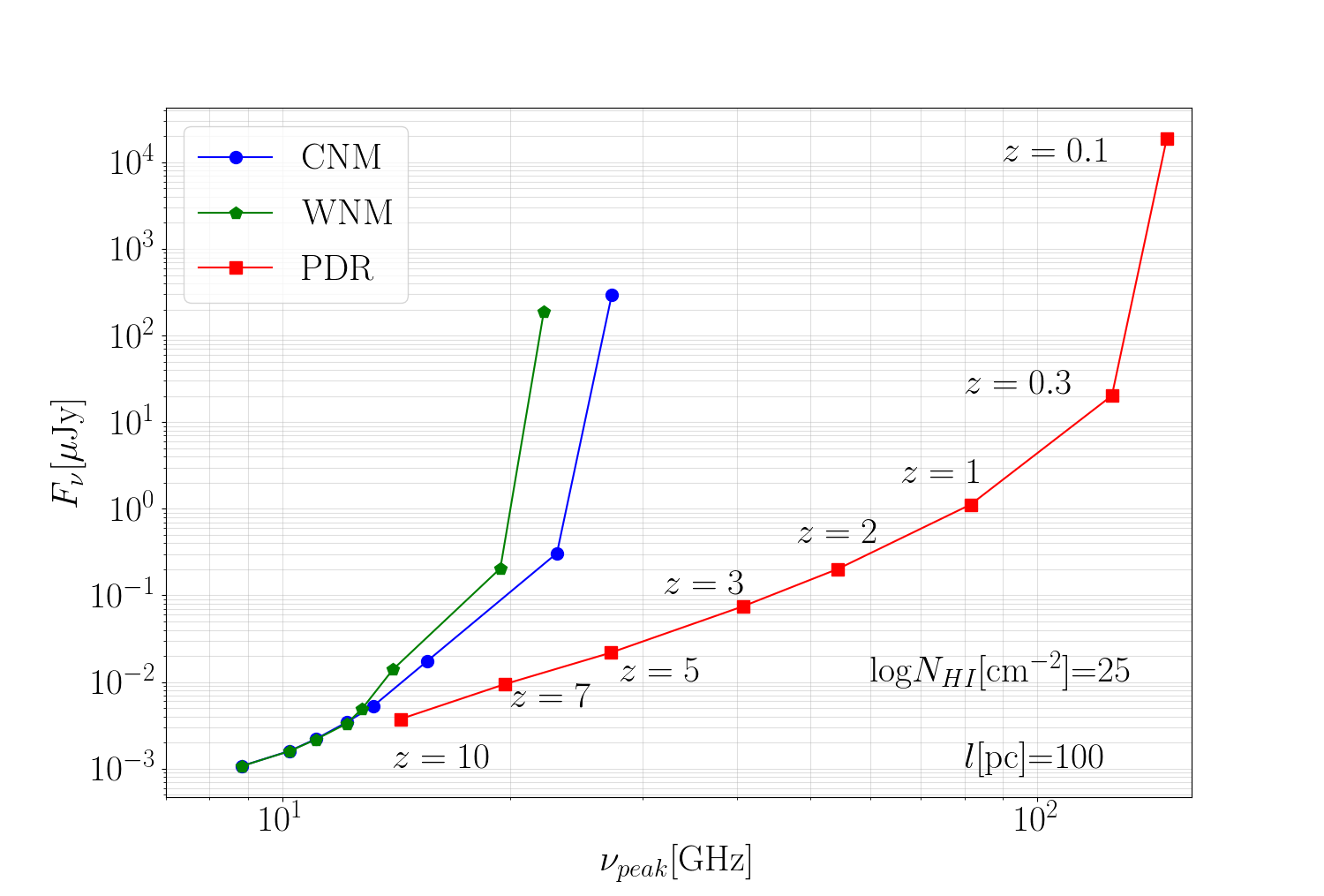}
\caption{The peak frequency and the corresponding flux density at the peak frequency for AME at different redshift. The flux density values scale with $l^2$ and $N_{\mbox{\tiny H}}$.}\label{fig:pflux}
\end{figure}

\begin{table*}
    \centering
    \begin{tabular}{|l|c|c|}\hline
        %\colhead{Parameter name [units]} $\vert$ & \colhead{JAGUAR models input} $\vert$ & \colhead{DSFG models input} \\\hline
        Parameter name [units] & JAGUAR models input & DSFG models input \\\hline   {\tt{tau\_main}} [Myr] & Binned average & $10\string^\{1.0,2.0,3.0,4.0\}$\\
        {\tt{age\_main}} [Myr] & Binned average & $10\string^\{2.35, 2.45, 1.5,2.0,2.5,3.0, 3.5\}$\\
        {\tt{metallicity}} [$Z$] & Nearest acceptable to binned average & \{0.004, 0.008, 0.02\}\\
        {\tt{Av\_ISM}} [mag] & \{Binned average, 0.25, 0.5, 1, 2, 4, 8\} & \{1.9, 3.6, 3.9, 4.1, 4.2, 1, 2, 3, 4, 5, 6, 8\}\\
        {\tt{temperature}} [K] & $T_d(\{10,20,30,45,60,75,90\}\,$K$,z)$ & $T_d(\{10,20,30,45,60,75,90\}$K$,z)$\\
        {\tt{redshift}} [z] & Binned average & Same as for JAGUAR models\\\hline
    \end{tabular}
    \caption{The parameter values used as input to CIGALE. ``Binned average'' refers to the average of each bin defined by the range of stellar mass and redshift, and was calculated for the JAGUAR models. $T_d(T_*,z)$ refers to the Eq. \ref{eq:at} in \ref{subsec:radsed}. For the DSFG model set, the first two values for {\tt{age\_main}} and the first five values for {\tt{Av\_ISM}} were taken directly from the two studies discussed in Section \ref{subsec:DSFGs}. Notice that although we created an ensemble of DSFG SEDs based on these parameters, in this paper, we show the result based on a set of specific parameters: {\tt{tau\_main}} = 100 Myr, {\tt{age\_main}} = 100 Myr, {\tt{metallicity}} = 0.02 and {\tt{Av\_ISM}} = 4}
    \label{tab:params}
\end{table*}

\section{Model galaxy populations} \label{sec:galpop}
To investigate the capability of radio telescope to detect high-$z$ galaxies, one needs to simulate SEDs of realistic galaxy populations that we observe. Recently JWST discovered high-$z$ galaxies which are the rest-frame UV-bright star-forming galaxies with star formation rate of 1--10 M$_{\odot}$ yr$^{-1}$\citep[e.g.,][]{atek_etal_2023b,borsani_etal_2023,franco_etal_2024,robertson_etal_2023,castellano_etal_2024,wang_etal_2023,morishita_etal_2024}. However, we also observe dusty star-forming galaxies that are not visible in the rest-frame UV but bright in the rest-frame FIR. In this study, we adopt two different galaxy populations in our Universe and model their SED based on the average stellar population parameters sampled from each galaxy population. In this section, we describe the galaxy population that was used to create an ensemble of model model SEDs. 

\subsection{UV-bright star-forming galaxies}\label{subsec:JAGUAR}
We use the simulated galaxy catalog, JAGUAR \citep{Williams_2018}, created for the JWST Advanced Deep Extragalactic Survey \citep[JADES,][]{jades_2023}. The JAGUAR catalog\footnote{\url{https://fenrir.as.arizona.edu/jaguar/index.html}} is a compilation of the star-forming galaxy models in the 10 different realizations. The catalog is split into star-forming and quiescent galaxies --- only the star-forming galaxies in each of the 10 different realizations were used. The entire compilation was used in order to improve the number counts of higher-mass galaxies at high redshift. Each realization was created by running the JAGUAR code, which is described in detail in \citet{Williams_2018}. Essentially, the JAGUAR code selects galaxies from a parent population that was generated by models designed to match empirical observations (stellar mass and UV luminosity function) at low redshift and to smoothly extrapolate physical properties to higher redshifts. From here on, any reference to the JAGUAR mock catalog refers to the complete compilation of the 10 realizations of star-forming galaxies.

In order to analyze this data set, the galaxies were first binned into ten different redshift ranges of equal comoving volume, beginning at $z=0$ and ending at $z=15$. The redshift bins are listed in the first row of Table \ref{tab:counts} in Appendix. Second, each of these galaxies was assigned to one of the following mass bins (denote log$_{10}(M_*/M_{\odot})$ by $\mathcal{M}$): $6\leq\mathcal{M}<7$, $7\leq\mathcal{M}<8$, $8\leq\mathcal{M}<9$, $9\leq\mathcal{M}<10$, $10\leq\mathcal{M}<11$, and $11\leq\mathcal{M}$. Because the number counts for the $11\leq\mathcal{M}$ bins were often small, these galaxies are not included in the results presented later in this paper. We also investigate the galaxies in a ``characteristic mass'' bin. In other words, these galaxies had to be within 0.5 dex of the characteristic mass at the mean redshift of their redshift bin. Quantitatively, these galaxies satisfied $z_{i-1}< z\leq z_{i}$ and $|\mathcal{M}-M_{2,M}^*(\bar{z}_{i})|<0.5$ for some $i\in\{1,...,10\}$, where $\bar{z}_{i}$ is the mean redshift of the galaxies in the redshift bin $[z_{i-1},z_i)$ and $M_{2,M}^*(z)=10.30-0.15z$ is the characteristic mass of M$_{*}$ galaxy described in \citet[][see Sec 3.1, Eq. 8]{Williams_2018}. The numbers of galaxies in each stellar mass and redshift bin, are summarized in Table~\ref{tab:counts}.

For higher masses, some moderate-to-high redshift bins contained few galaxies or none at all. And the JAGUAR catalog has no galaxies past $z=15$. In order to produce models at these higher redshifts, a linear fit was performed for each parameter as a function of the time (i.e. the age of the universe, not redshift). These fits are plotted in Figure \ref{fig:JAGUARfits}. Fitting as a function of time was deemed preferable to fitting as a function of redshift for the following reason: the only parameter with a noticeable correlation with time/redshift, {\tt{age\_main}}, appears to be non-linear as a function of redshift, while it is bounded by a linear function of time, the age of the universe. Additionally, it should be noted that the linear fit with respect to time, when plotted as a function of redshift (as in Figure \ref{fig:JAGUARfits}), flatten out at higher redshifts; this means that the fits, for $10<z<20$, are nearly constant. So the model galaxy properties at the highest redshifts are similar to those at intermediate redshifts ($z\sim10$). Additionally, instead of evaluating the fits at high-redshift bins with comoving volumes equal to that of the bins for $z<15$, which would have yielded only two extra bins, it was decided to do so for bins with half the comoving volume, yielding four extra bins.

Lastly, the criterion for which redshift bins to use extrapolated parameter values was that the given mass and redshift bin have 2 or fewer galaxies: this was based on the empirical observation that using the median values of bins with two or fewer galaxies yielded parameter values that deviated significantly from those of galaxies with similar masses and redshifts. Thus, the $\mathcal{M}_{char}$, $6\leq\mathcal{M}<7$ and $7\leq\mathcal{M}<8$ bins used median values all the way up to $z\sim13.2$, after which extrapolated values were used; the $8\leq\mathcal{M}<9$ bin used median values up to $z\sim11.1$; the $9\leq\mathcal{M}<10$ bin used median values up to $z\sim9.3$; and the $10\leq\mathcal{M}<11$ bin used median values up to $z\sim5.2$. The input values for \texttt{CIGALE} used to create SEDs of galaxies from JAGUAR catalog are presented in Table~\ref{tab:params}.  

\subsection{Dusty star-forming galaxies} \label{subsec:DSFGs}

To create models representing DSFGs, two samples of DSFGs in literature were considered, and the average values of the physical parameters reported for these samples were used to create a subset of models with these exact values.

The first sample, presented in \citet{Ma_2019}, consists of \textit{Herschel}-selected DSFGs with follow-up \textit{Spitzer} observations and other ancillary data from, e.g. ALMA and SCUBA2. There are a total of 63 galaxies in the catalog, some of which were further classified as ``unlensed and ultrared'' DSFGS (48 galaxies) and ``unlensed, ultrared, and single-component" DSFGs (31 galaxies). Table 5 of \cite{Ma_2019} presents the averages of the physical properties derived from SED fitting with the {\tt{MAGPHYS}} program (they allowed the photometric redshift to be a free parameter that {\tt{MAGPHYS}} could constrain). The values reported for $A_{V}$ and the age of the stellar population were used as input parameters for {\tt{CIGALE}}; \citet{Ma_2019} report that $T_{dust}$ is poorly constrained by the data, so those values were not used --- instead, we use a range of values (10--90K) for the dust temperature at $z=0$ ($T_{d,*}$ in Equation~\ref{eq:tdust}) for both the JAGUAR catalog and DSFGs. Stellar mass is not an input parameter for {\tt{CIGALE}} --- rather, the default setting is to normalize the SFH to integrate to 1 $M_{\odot}$ --- so the stellar mass is used later as a multiplicative factor when analyzing and plotting the SEDs. The dust luminosity and SFR are also used after the SEDs have been produced to check that the models are indeed able to produce the high luminosities and SFRs reported for DSFGs.
%--- {\tt{CIGALE}} automatically outputs the SFR integrated over the last 10 Myr (denoted $SFR_{10Myr}$), making the comparison straightforward. As an example, when models produced low $SFR_{10Myr}$ values, galaxies with {\tt{tau\_main}} = 10 Myr and {\tt{age\_main}} = $10^{3.5}$ Myr resulted in models with $SFR_{10Myr}~1$ for a $10^{11}M_{\odot}$ galaxy, which is well below the $10^{1}-10^{3}$ range expected for DSFGs. (While some model dust luminosities were on the low side, they were never as far from realistic as the $SFR_{10Myr}$ values.)

The second sample was presented and analyzed in \citet{Talia_2021}. This sample consists of 197 radio-selected, UV-dark galaxies identified using the VLA-COSMOS 3 GHz Large Project. Within the total sample, the authors carried out multiwavelength analysis of a ``primary'' subsample with FIR and MIR/NIR data. They present the medians of the physical properties of the total sample and primary subsample in their Tables 1 and 2, respectively. Again, the $A_V$ value is used directly as an input parameter, while the $T_{dust}$ is ignored in favor of the previously identified range of values. The stellar mass will be used in the analysis in the same way as before. And the infrared luminosities and SFR could again be used to check for consistency with the observed values.

In addition to the reported values for $A_V$ and stellar age, regularly spaced sets of values in what were deemed reasonable and interesting ranges were also used as input values for these parameters. These input values were: {\tt{Av\_ISM}}=\{1,2,3,4,5,6,8\} and {\tt{age\_main}}=10$\string^$\{1.5,2.0,2.5,3.0,3.5\}\footnote{The upper two values of {\tt{age\_main}} are incompatible with the age of the universe at the highest redshifts. For such incompatible input values {\tt{CIGALE}} will still output an SED, but these are not used in the analysis.}. Two other {\tt{CIGALE}} parameters, {\tt{tau\_main}} and {\tt{metallicity}}, were likewise given a set of regularly spaced values: 10$\string^$\{1.0,2.0,3.0,4.0\} and \{0.004, 0.008, 0.02\}, respectively. The set of redshifts used for the DSFG models was the same as that used for the JAGUAR-derived models. The input values for \texttt{CIGALE} used to create SEDs of DSFG are presented in Table~\ref{tab:params}.  
%%%%%%%%%%%%%%
We note that more recent analysis of a larger statistical sample of DSFGs based on JWST and ALMA\citep{mckinney_etal_2025} and COSMOS2020 multiwavelength catalog\citep{gentile_etal_2024} also present the similar distributions of the extinction values and the stellar masses.  
%%%%%%%%%%%%%%%

To summarize, the distribution of the parameters encountered in the literature search was used to inform the range of evenly spaced input values for {\tt{CIGALE}}; in addition to these regularly spaced values, a few specific values coming from the two samples above were used as input for the appropriate parameter. For the sake of clarity, however, Figures \ref{fig:compDSFG_lowf} and ~\ref{fig:compDSFG_highf} only display models with a small selection of parameter values. In particular, while {\tt{redshift}} and {\tt{temperature}} vary across the model SED's displayed in these figures, a fixed set of value for {\tt{tau\_main}}, {\tt{age\_main}}, {\tt{Av\_ISM}} and {\tt{metallicity}} was used for each model: {\tt{tau\_main}} = 100 Myr, {\tt{age\_main}} = 100 Myr, {\tt{metallicity}} = 0.02 and {\tt{Av\_ISM}} = 4, which are not widely different from the values in the literature.

\begin{table*}
\footnotesize
%\begin{rotatetable*}
\centering
    \begin{tabular}{|c|c|c|c|c|c|c|c|c|c|c|c|c|}\hline
        telescope & 110MHz & 300MHz & 770MHz & 1.4GHz & 6.7GHz & 12.5GHz & 27GHz & 41GHz & 93GHz & 150GHz & 250GHz & 350GHz \\
        name & $\mu$Jy & $\mu$Jy & $\mu$Jy & $\mu$Jy & $\mu$Jy & $\mu$Jy & $\mu$Jy & $\mu$Jy & $\mu$Jy & $\mu$Jy & $\mu$Jy & $\mu$Jy\\\hline
        SKA & 26 & 14 & 4.4 & 2.0 & 1.3 & 1.2 & -- & -- & -- & -- & -- & -- \\
        ngVLA & -- & -- & -- & 0.24 & 0.14 & 0.16 & 0.17 & 0.21 & 0.40 & -- & -- & -- \\
        ALMA & -- & -- & -- & -- & -- & -- & -- & -- & 11 & 12 & 20 & 26 \\\hline
    \end{tabular}
    \caption{The rms/1-$\sigma$ sensitivity values for each facility, for an integration time of 1 hour. ALMA sensitivity is estimated using ALMA Observing Tool with the 12m array for 1$''$ angular resolution at a sky position of (RA,Dec)=($0^\circ$,$0^\circ$). For ngVLA and SKA, the predicted sensitivity values are quoted: for ngVLA, the naturally weighted sensitivity values are used from the project website; for the SKA, the sensitivity values of SKA-Mid AA4 baseline design array \citep{braun_v1.0} for $\approx 1''$ beam were used for the sensitivity calculator.}
    \label{tab:rms}
%\end{rotatetable*}
\end{table*}

\section{Results} \label{sec:result}

The main motivation of this study is to provide a prospect of radio-millimeter observation to detect galaxies in the high-$z$ Universe, using the current and the next-generation radio--millimeter wavelength telescopes (ALMA, ngVLA and SKA). In this section, we present the variation of the galaxy SED with redshift up to $z\sim20$, focusing on the radio--FIR wavelength and based on the simulation of galaxy SEDs using realistic stellar population model parameters with the simple prescriptions of emission processes and the CMB effect as described in Section~\ref{sec:sim}. 

In Section~\ref{subsec:amecont}, we present our investigation into the significance of AME contribution to the radio--FIR SED. Then we illustrate the effect of the CMB and assumed dust temperature on the radio--FIR SED in Section~\ref{subsec:var}. Finally, in Section~\ref{subsec:simul}, we present the flux density variation against the sensitivities of the telescopes at a set of key observing frequencies for SKA, ngVLA and ALMA (see Table~\ref{tab:rms}) as a function of redshift for the simulated galaxy populations (i.e., UV-bright star forming galaxies and dusty star forming galaxies). We note that although we compare the predicted flux densities to the sensitivities specific to SKA, ngVLA, and ALMA, our results are generally applicable to any radio-FIR frequencies. 

\subsection{Contribution of AME to the radio SED}\label{subsec:amecont}

To compute the flux density of AME from ISM with different redshifts, we first compute the AME emissivity for a range of redshifts ($0.1<z<10$) with an enhanced ambient radiation field driven by the evolution of CMB energy density (Equation~\ref{eq:ambrad}), for different ISM conditions: CNM, WNM, and PDR. Then, we compute the AME SED using Equation~\ref{eq:ameflux} with the CMB-enhanced AME emissivity for an ISM with a 100 pc diameter and a $10^{25}$ cm$^{-2}$ column density. Here we note that the column density is purposely chosen to be high such that we can demonstrate that the contribution from AME to the radio SED is not significant even if we assumed an extreme ISM environment (\Nh$\sim10^{25}$cm$^{-2}$) for strong AME.

In Figure~\ref{fig:pflux}, we show the variation of the peak flux density of AME at the peak frequency in different redshifts ($0.1<z<10$) for CNM, WNM, and PDR. With increasing redshift, the peak flux decreases and the peak frequency moves to a lower frequency. The peak flux density of AME from CNM and WNM is already below 0.1 $\mu$Jy if $z>0.3$ while the AME flux from PDR can be similarly bright (0.1 $\mu$Jy) even in $z\sim2$. Note that the flux density values scale with $l^2$ and $N_{\mbox{\tiny H}}$. If we consider a more normal ISM condition with column density less than the currently assumed value ($<10^{25}$cm$^{-3}$), the AME will be even fainter than what we estimate in this study. One possible way of boosting AME flux density is to increase the AME emitting region size $l$ in Equation~\ref{eq:ameflux} (a factor 10 decrease in the column density is compensated by 3 times larger AME region size, $l=300$pc), however assuming large AME region (a few hundreds parsec), larger than the currently assumed value $l=100$ pc, is unrealistic as well. Therefore, we conclude that the contribution of AME to the radio SED with continuum sensitivity larger than a level of $\mu$Jy is insignificant for galaxies with $z\gtrsim1$. 

\subsection{Illustration of the impact of the CMB and dust temperature to the radio--FIR SED} \label{subsec:var}
In this section, we present the effect of the CMB and dust temperature on the radio--FIR SED of galaxies.

\subsubsection[Impact of CMB for high-redshift galaxy SED]{Impact of CMB for high-$z$ galaxy SED} \label{subsubsec:zcmb}
\begin{figure*}
    \centering
    \includegraphics[width=6in]{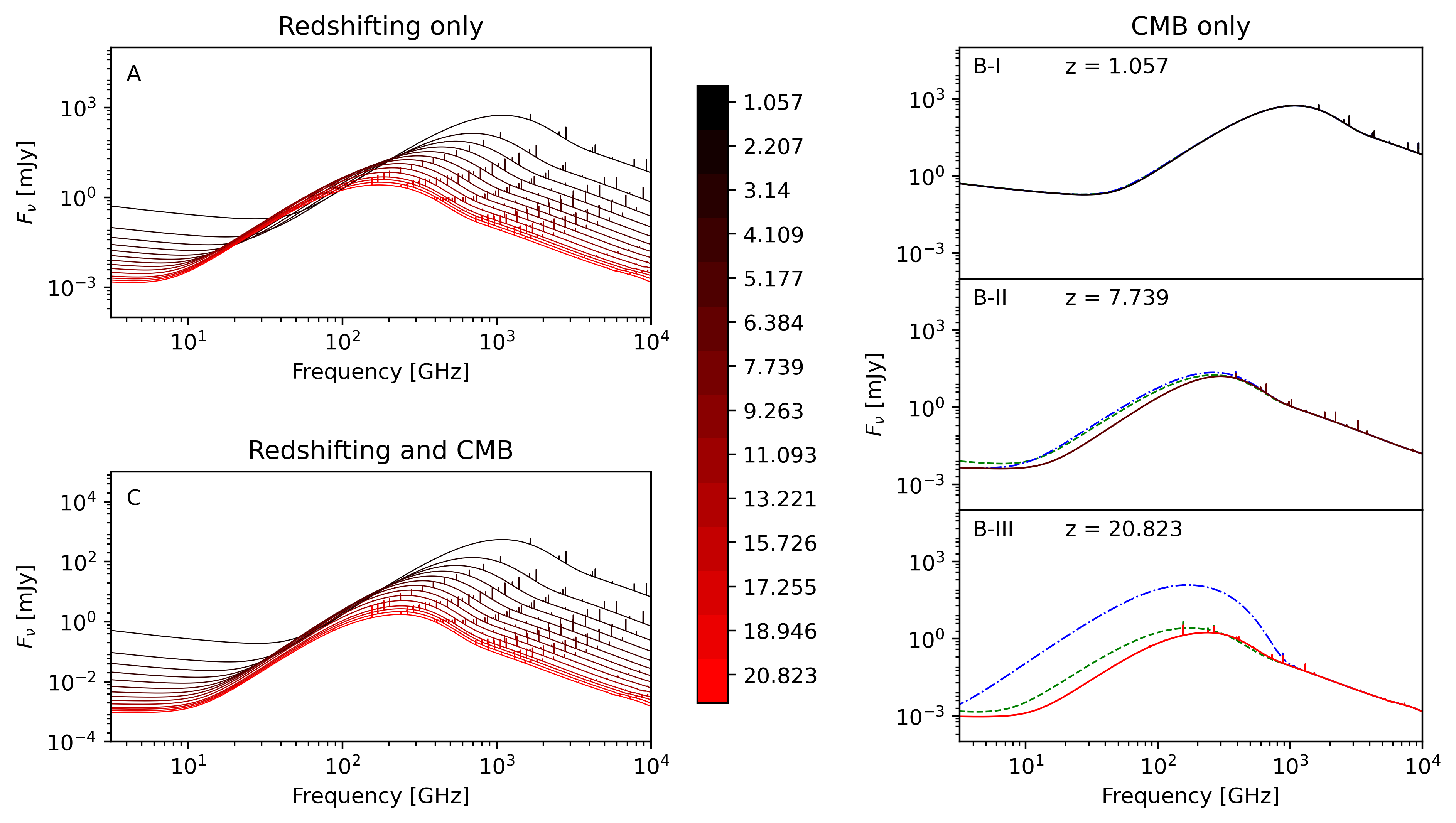}
    \caption{Panel A illustrates the variation with the redshift of 14 SEDs without including the CMB heating and contrast, for models with all other parameters the same: {\tt{tau\_main}} = 100 Myr, {\tt{age\_main}} = 100 Myr, {\tt{metallicity}} = 0.02, {\tt{Av\_ISM}} = 4, and $T_{d,*}$ = 30 K (note that $T_d$ does vary with redshift). Panels B-I, B-II, and B-III show the effect of the CMB for three of the same models at different redshifts: $z = 1.057, 7.739, 20.823$. The green dashed lines show the SED before correcting for the CMB heating and contrast; the dash-dotted blue lines show the SED after adding the additional IR luminosity due to CMB heating but not subtracting the modified black body function for CMB as a background; the red solid lines show the SED after correcting for the additional IR luminosity and the background. Panel C demonstrates the redshift variation of the model SEDs in Panel A by incorporating the proper correction of the CMB effect. A stellar mass of $10^{11}M_{\odot}$ was used to scale the SEDs.}
    \label{fig:zcmb}
\end{figure*}

The CMB changes the shape of galaxy SED in radio via synchrotron suppression (Section~\ref{subsec:radsed}) and in FIR via dust heating and decreasing contrast (Section~\ref{subsec:firsed}). Figure~\ref{fig:zcmb} shows the DSFG SEDs for 14 different redshift bins ($1\lesssim z \lesssim20$) with every input SED parameter including dust temperature ($T_{d,*}$) fixed. In panel A of Figure~\ref{fig:zcmb}, we show the variation of galaxy SED with the redshift, without including the CMB effect. Because of the negative $K$-correction, the FIR SED flux at a given observing frequency does not decrease, and even increases with redshift for certain frequencies. However, if we include the CMB in the SED model, as illustrated in panel B of Figure~\ref{fig:zcmb}, the initial SED (green dashed line) is boosted by CMB heating (blue dot-dashed line) but the observed contrast is decreased by the higher thermal background due to the CMB (red solid line). The difference between the initial SED (green dashed line) and the final SED (red solid line) becomes large in high-redshift for fixed $T_{d,*}$. After including the CMB properly, panel C in Figure~\ref{fig:zcmb} shows the variation of galaxy SEDs with redshift. The FIR SED after the `CMB corrected' negative $K$-correction (panel C) is what we use in this study. 

\subsubsection[Impact of dust temperature for high-redshift galaxy SED]{Impact of dust temperature for high-$z$ galaxy SED} \label{subsubsec:temp}

\begin{figure*}
    \centering
    \includegraphics[width=6in]{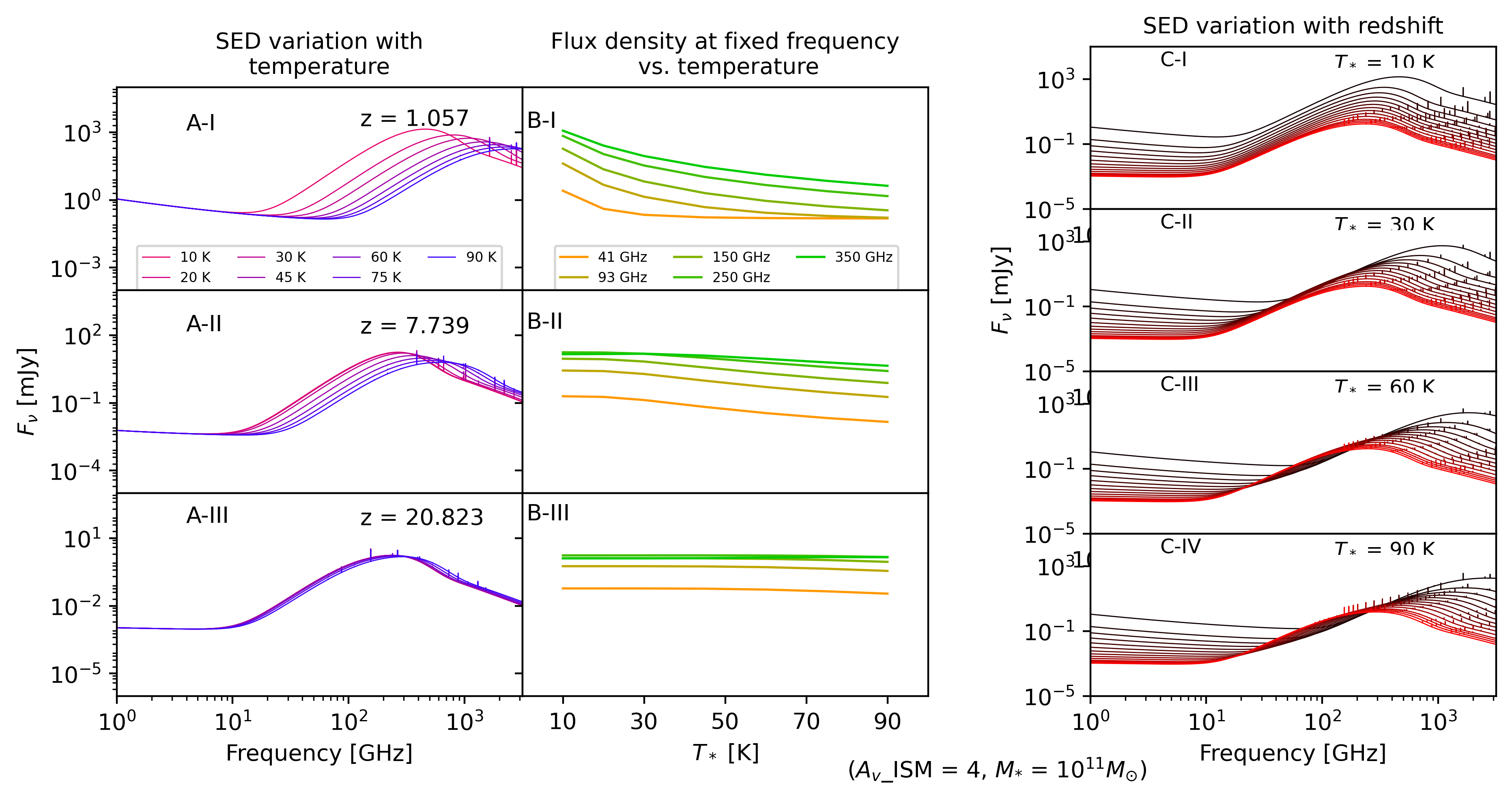}
    \caption{Panels A-I, A-II, and A-III illustrate how SEDs vary with temperature at three redshifts $z = 1.057, 7.739, 20.823$. All physical parameters, except for temperature, are the same as for the SEDs in Figure \ref{fig:zcmb}. Panels B-I, B-II, and B-III show the corresponding flux densities at five selected frequencies: 41, 93, 150, 250 and 350 GHz. Panels C-I though C-IV illustrate the variation of SEDs with redshift, as in Panel C of Figure \ref{fig:zcmb}, but now for four different $T_{d,*}$'s (Panel C-II is the same as Panel C in Figure \ref{fig:zcmb}).}
    \label{fig:temp}
\end{figure*}

For a fixed IR luminosity, dust temperature changes the FIR SED shape by shifting the FIR peak frequency (a lower dust temperature moves the frequency of the FIR SED peak to a lower frequency). In Figure \ref{fig:temp}, we show the variation of SEDs with the assumed dust temperature parameter ($T_{d,*}=$10--90 K). One may observe that in Panel A-{\sc{I}}, where the CMB correction is negligible in low $z$, the FIR SED shape changes significantly with different dust temperatures. With the FIR SED peak shifted to higher frequencies as the dust temperature increases, the end result is that the flux density at all considered frequencies (41, 93, 150, 250, 350 GHz) decreases with temperature as shown in Panel B-{\sc{I}}. At higher redshifts with increasing CMB temperature, however, the CMB correction has a significant impact on the FIR SED with low dust temperature, while the impact is less significant for high dust temperature. As a result, the difference in shape between the FIR SED of low and high dust temperature becomes less noticeable with increasing redshift and the SEDs are all converging at high redshift even though the inherent dust temperatures are different, as shown in Panel A-{\sc{II}} and A-{\sc{III}}. 

\subsection{Simulation of the radio--FIR SED observation using SKA, ngVLA and ALMA}\label{subsec:simul}
In this section, we present the variation of the radio-FIR flux densities as a function of redshift, at several different observing frequencies for SKA, ngVLA, and ALMA and compare them to each telescope's sensitivity values. For ALMA, we use ALMA Observing Tool\footnote{\url{https://almascience.nrao.edu/proposing/observing-tool}}. For SKA, we adopt the predicted values from SKA Sensitivity Calculator\footnote{\url{https://sensitivity-calculator.skao.int/}}, and for ngVLA, we adopt the predictecd values from the project website\footnote{\url{https://ngvla.nrao.edu/page/performance}}.  

\subsubsection{UV-bright star-forming galaxies}\label{subsubsec:jaguargal}
\begin{figure*}
    \centering
    \includegraphics{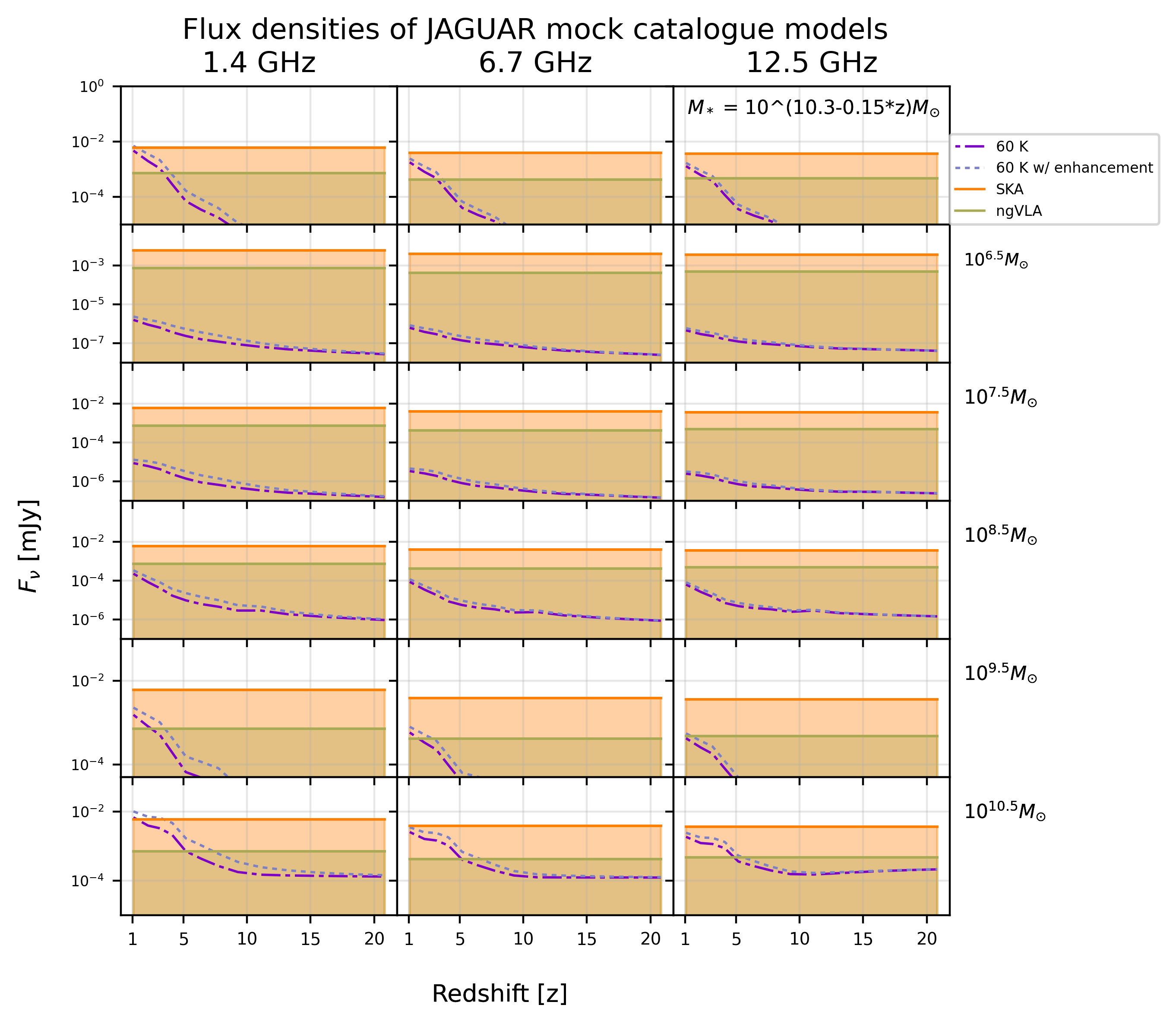}
    \caption{The radio flux density variation as a function of redshift for a given observing frequency, for galaxies sampled from the JAGUAR catalog. Each row corresponds to a different mass bin --- from top to bottom: $M_{char}(z)$, $6\leq\mathcal{M}<7$, $7\leq\mathcal{M}<8$, $8\leq\mathcal{M}<9$, $9\leq\mathcal{M}<10$, and $10\leq\mathcal{M}<11$. The $11\leq\mathcal{M}$ is not plotted because it appears to be unreliable due to small sample sizes. For a given mass bin, the flux density of a model with the mass given by the upper right corner of the right-most column is plotted vs. redshift and compared to the $3\sigma$ sensitivity of SKA and ngVLA for 1 hour integration based on the 1$\sigma$ sensitivity in Table~\ref{tab:rms}. The dot-dashed line is the radio SED model with the synchrotron dimming only (`version 1') and the dotted line is the radio SED model with the synchrotron dimming and magnetic field enhancement (`version 2'), for 60K dust temperature. The flux densities for the frequencies lower than 1.4 GHz (i.e., 0.11, 0.3, 0.77 GHz) are not shown because no UV star-forming galaxies in our model SEDs are detectable in those frequencies for the given sensitivities of the telescopes. Note that although the flux densities are shown based on the SED model with a 60K dust, the effect of dust temperature is negligible for the low-frequency radio SED and the behavior of flux density in each panel is same for any dust temperature.}
    \label{fig:compJAGUAR_lowf}
\end{figure*}

\begin{figure*}
    \centering
    \includegraphics{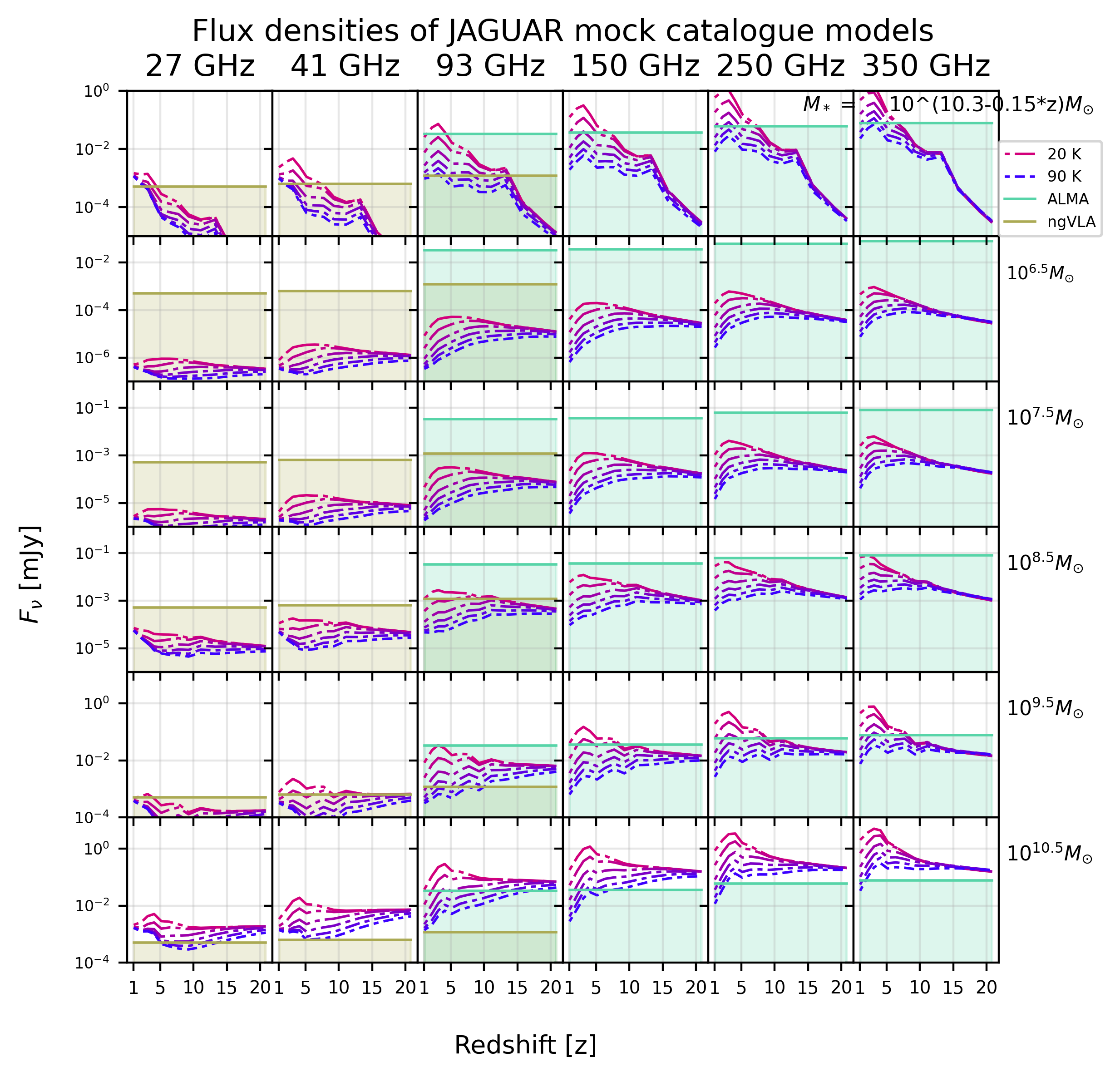}
    \caption{Same as Figure~\ref{fig:compJAGUAR_lowf} but for higher observing frequencies. ALMA $3\sigma$ sensitivities are presented for high-frequency observing bands. Overall, when ngVLA operates, $\approx 100$ GHz frequency is the most promising observing window for high-$z$ galaxy detection. The flux densities based on the model SEDs with different dust temperatures sequentially increasing from 20 to 90 K (20, 30, 45, 60, 75, 90 K) are shown by different lines and colors.}
    \label{fig:compJAGUAR_highf}
\end{figure*}

Figures~\ref{fig:compJAGUAR_lowf} and \ref{fig:compJAGUAR_highf} show the redshift variation of the flux density of radio--FIR SEDs from star-forming galaxies in the JAGUAR catalog (Section~\ref{subsec:JAGUAR}) for selected observing frequencies available in SKA, ngVLA, and ALMA split by low-frequency group (0.11, 0.30, 0.77, 1.4, 6.7, 12.5 GHz) and high-frequency group (27, 41, 93, 150, 250, 350 GHz). Figure~\ref{fig:compJAGUAR_lowf} and Figure~\ref{fig:compJAGUAR_highf} have multiple panels showing the flux as a function of redshift ($1\lesssim z \lesssim 20$) for 60 K temperature dust (Figures \ref{fig:compJAGUAR_lowf} for low frequencies) and for a range of dust temperatures with 20 K$\leq T_{d,*} \leq$90 K (Figure~\ref{fig:compJAGUAR_highf} for high frequencies). The panels in each column show the result for different observing frequencies. The panels in each row show the result for galaxies selected in different stellar mass bins: the top row presents the galaxies with characteristics stellar mass (i.e., M$^{*}$, stellar mass at the knee of galaxy stellar mass function) at redshift $z$ parameterized in \cite{Williams_2018} and the other rows beginning with the second row present the galaxies with stellar mass bins (increasing mass with descending row). The binned average of the {\tt{Av\_ISM}} values are used for each mass/redshift bin --- {\tt{Av\_ISM}} thus varies between mass bins and across redshifts. Note that in Figure~\ref{fig:compJAGUAR_highf} for high frequencies, the flux densities with different dust temperatures are converging at high redshift ($z=20$) because the high-redshift CMB that is hotter than dust erases the difference in the SED shape due to different dust temperature as illustrated by Panel A in Figure~\ref{fig:temp}.

Figure \ref{fig:compJAGUAR_lowf} shows the case of low frequencies (1.4, 6.7, 12.5 GHz) available both in SKA and ngVLA. The flux densities are shown based on the SED model with a 60K dust because the effect of dust temperatures is negligible for the low-frequency radio SED and the behavior of flux density in each panel is same for any dust temperature. Note that the results for the lower frequency ($<1$ GHz) observing bands are not presented because the flux densities of UV star-forming galaxies in our model are lower than the telescope sensitivities. The shaded region with solid horizontal line in each panel indicates $3\sigma$ RMS sensitivity with 1-hour integration for SKA (peach color) and for ngVLA (olive color) at the chosen observing frequency. Figure \ref{fig:compJAGUAR_highf} shows the case of high frequencies available in ngVLA (27, 41, 93 GHz) and in ALMA (93, 150, 250, 350 GHz). The flux densities based on the model SEDs with different dust temperatures sequentially increasing from 20 to 90 K (20, 30, 45, 60, 75, 90 K) are shown by different lines and colors. The shaded region with solid horizontal line in each panel indicates $3\sigma$ RMS sensitivity with 1-hour integration for ngVLA (olive color) and for ALMA (bright green color) at the chosen observing frequency.

In low-frequency ($\nu \lesssim 15$ GHz) observation (Figure~\ref{fig:compJAGUAR_lowf}), UV-bright star-forming galaxies sampled from the JAGUAR catalog are not detectable for all range of galaxy stellar masses M$_* < 10^{10.5}$M$_{\odot}$ and dust temperatures ($T_{d,*} \lesssim 90$) we are probing. This simply implies that the `average' star formation rate of such galaxies (SFR$\sim 1-10$ M$_{\odot}$ yr$^{-1}$) based on the SFR-M$_{*}$ relation in high-$z$ Universe \citep[e.g.,][]{iyer_etal_2018} is not strong enough to create a detectable radio flux based on the empirical correlation between radio and IR luminosity well established in the local universe. 

However, in high-frequency ($\nu \gtrsim 30$ GHz) observation (Figure~\ref{fig:compJAGUAR_highf}), the situation is promising. For $M^*$ galaxies, if the dust temperature is low ($T_{d,*} \lesssim 30$K), observations in 27 and 41GHz (covered by ngVLA) and in 150, 250 and 350GHz (covered by ALMA) can detect the galaxies up to $z\sim5$. At 93GHz, ngVLA, with $\sim 30\times$ higher sensitivity than ALMA (see Table~\ref{tab:rms}), can detect M$^{*}$ galaxies  with a reasonable dust temperature (20--60K) up to $z\sim10$ and beyond. Lower stellar mass galaxies (M$_* < 10^9$M$_{\odot}$) are still too faint to be detected by ngVLA and ALMA. However, the massive galaxies (M$_* > 10^{10}$M$_{\odot}$) can be detected in any redshift. In particular, when ngVLA operates, 93GHz is the most sensitive window to probe galaxies independent of redshift even if the galaxy mass is not extreme ($10^{9.5} <$M$_{*}/$M$_{\odot} <10^{10.5}$).  

One caveat for interpreting the result based on simulated galaxy SEDs is that the assumed dust temperature should be realistic such that the inferred dust mass should take up a reasonably small fraction of stellar mass: for low dust temperature ($\lesssim 20$ K), the derived dust mass can be as unrealistically large as $10^{10}$M$_{\odot}$. Therefore we checked the inferred dust mass from the galaxy SED for each redshift bin in Figure~\ref{fig:compJAGUAR_highf} and verified that the dust mass is always less than the galaxy stellar mass for every stellar mass bin. 
%except for the case where the assumed dust temperature is very low: $T_{d,*}=10$ K. 
This implies that the UV-bright galaxies are unlikely to have a dust with very low ($\sim10-20$ K) temperature which is not visible anyway in high-redshift ($z\gtrsim10$) due to the high temperture CMB. 

\subsubsection{Dusty star-forming galaxies}\label{subsubsec:DSFG} 
\begin{figure*}
    \centering
    \includegraphics{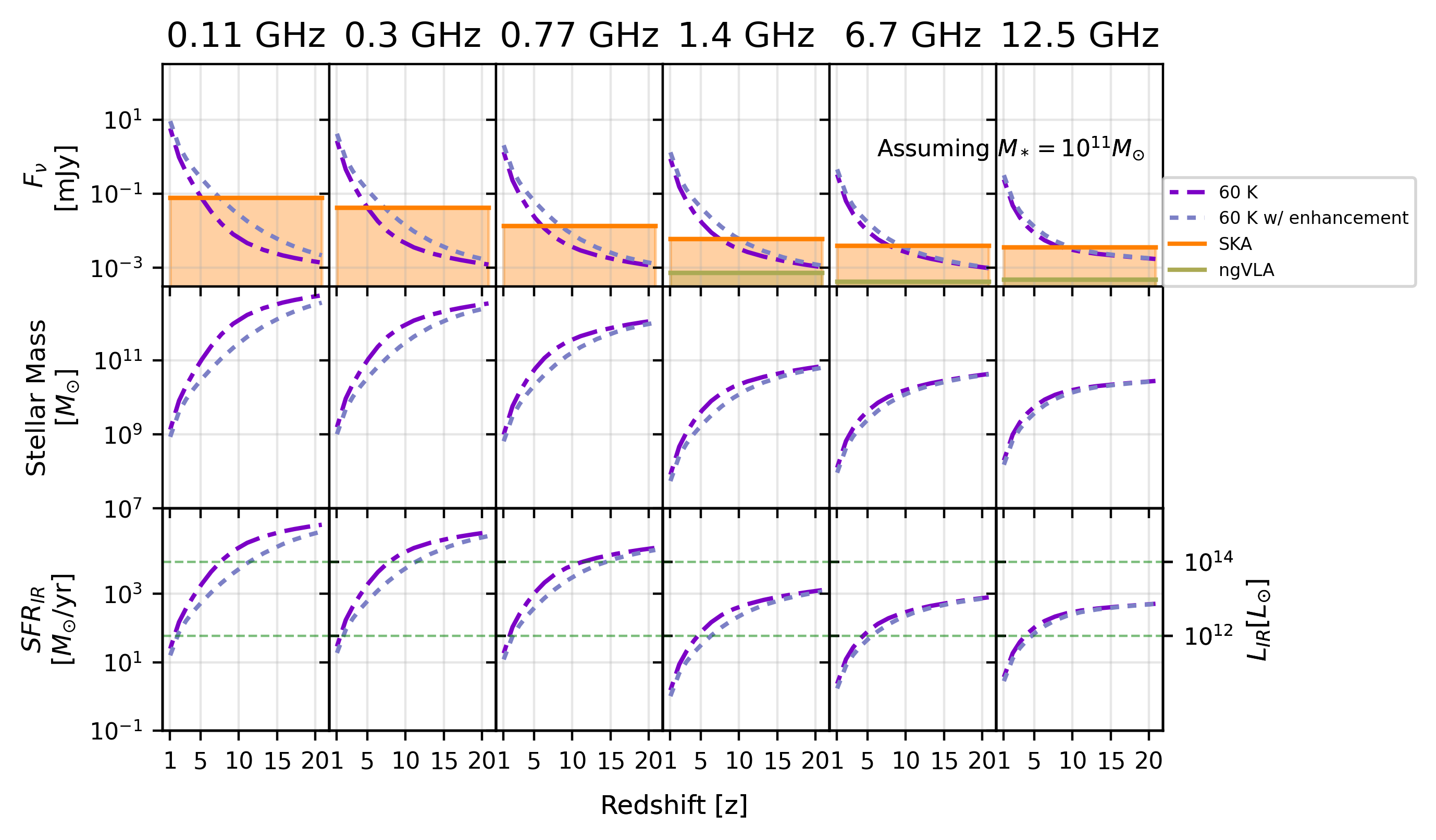}
    \caption{In the top row, the flux density of a DSFG with $M_*=10^{11}M_{\odot}$ is plotted as a function of redshift and compared to the $3\sigma$ RMS of the SKA and/or ngVLA (Table~\ref{tab:rms}), depending on the observing frequency band and sensitivity (i.e., lower RMS sensitivity is used when both SKA and ngVLA have the same observing frequency. The second row illustrates the minimum stellar mass needed to scale the normalized SED such that its flux density is equal to $3\sigma$ RMS of one of the facilities (the SKA's sensitivity is used for 0.11, 0.3, 0.77 GHz and ngVLA's for 1.4, 6.7, and 12.5 GHz). The third row presents the equivalent star formation rate averaged over the last 10 Myrs, $SFR_{10Myr}$ from \texttt{CIGALE}. The right-most y-axis of this row gives the equivalent IR luminosity, for this specific model. Like Figure~\ref{fig:compJAGUAR_lowf}, the dot-dashed line is the radio SED model with the synchrotron dimming only (`version 1') and the dotted line is the radio SED model with the synchrotron dimming and magnetic field enhancement (`version 2') for 60K dust temperature. Note that although the derived quantities in each panel are shown based on the SED model with a 60K dust, the effect of dust temperature is negligible for the low-frequency radio SED and the behavior of the quantities in each panel is same for any dust temperature.}
    \label{fig:compDSFG_lowf}
\end{figure*}
\begin{figure*}
    \centering
    \includegraphics{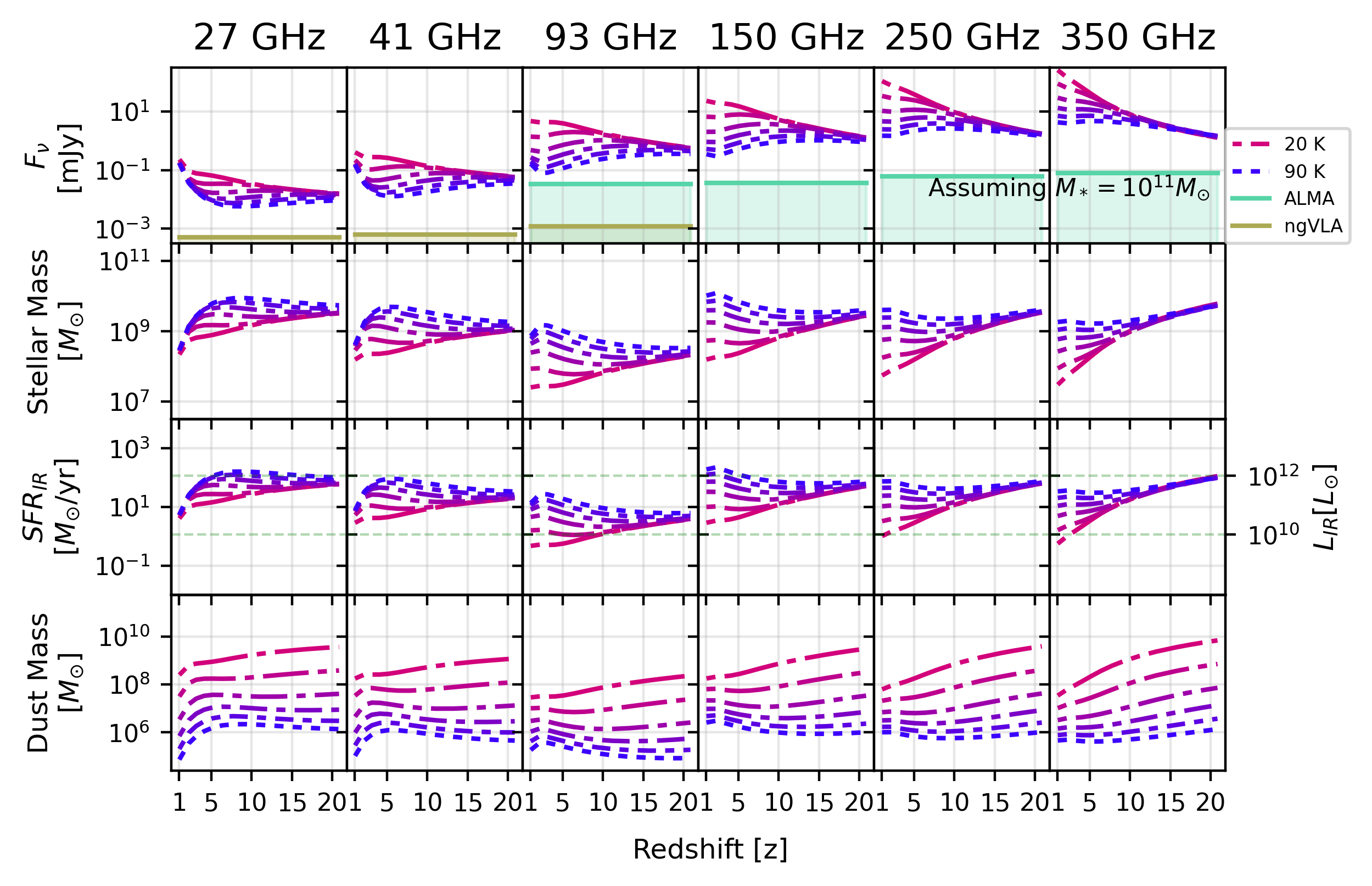}
    \caption{Same as Figure~\ref{fig:compDSFG_lowf} but for higher observing frequencies. ALMA $3\sigma$ sensitivities are presented for high-frequency observing bands. In addition, in the fourth row, the minimum dust masses for the galaxies to be detected with $>3\sigma$ for the fixed IR luminosity with different assumed dust temperatures, are shown as a function of redshift. The derived quantities based on the model SEDs with different dust temperatures sequentially increasing from 20 to 90 K (20, 30, 45, 60, 75, 90 K) are shown by different lines and colors.}
    \label{fig:compDSFG_highf}
\end{figure*}

Figures~\ref{fig:compDSFG_lowf} and \ref{fig:compDSFG_highf} show the redshift variation of the flux density of radio--FIR SEDs from dusty star-forming galaxies (Section~\ref{subsec:DSFGs}) for selected observing frequencies (low- and high-frequency group) available in SKA, ngVLA, and ALMA. Because of the same reason (i.e., negligible effect of dust temperature on the low frequency SED) for Figure~\ref{fig:compJAGUAR_lowf} and~\ref{fig:compJAGUAR_highf}, we show SEDs with a 60K dust for the low frequencies (Figure~\ref{fig:compDSFG_lowf}) and with a range of dust temperatures (20 K$\leq T_{d,*} \leq$90 K) for the high frequencies (Figure~\ref{fig:compDSFG_highf}). However, unlike the previous figures (Figure~\ref{fig:compJAGUAR_lowf} and~\ref{fig:compJAGUAR_highf}), the galaxy SEDs are not shown by the stellar mass bin. We create the SED for DSFGs with M$_{*}=10^{11}$M$_{\odot}$ using the average stellar population parameters and scale the SEDs by stellar mass if necessary. Here we emphasize that this model SED with M$_{*}=10^{11}$M$_{\odot}$ does not represent DSFGs at $z\sim 10$: such massive galaxies are unlikely to exist at $z\sim 10$. The reason for using $10^{11}$M$_{\odot}$ DSFG SED is to scale the model SED for different stellar masses and infer the minimal stellar mass and SFR that can be measured as function of redshift, as shown in Figure~\ref{fig:compDSFG_lowf} and~\ref{fig:compDSFG_highf}.

In Figures~\ref{fig:compDSFG_lowf} and~\ref{fig:compDSFG_highf}, the first row of each figure shows the flux density variation of DSFG with M$_{*}=10^{11}$M$_{\odot}$ for the chosen observing frequencies. Like Figure~\ref{fig:compJAGUAR_highf}, the flux densities with different dust temperatures are indistinguishable and converging at high redshift because of the high temperature CMB. To estimate the flux density for smaller stellar mass DSFGs, one can scale the SED with M$_{*}=10^{11}$M$_{\odot}$ by a factor of $\mbox{M}_{*}/10^{11}$M$_{\odot}$. The second row shows the minimum stellar mass of DSFGs that produces a flux density larger than the $3\sigma$ sensitivity limit as a function of redshift in each observing frequency for SKA, ngVLA and ALMA. Note that, for the overlapping frequencies: 1.4, 6.7, 12.5 GHz (available in both SKA and ngVLA) and 93 GHz (available in both ALMA and ngVLA), the value of $3\sigma$ sensitivity limit is adopted from the instrument with higher sensitivity (i.e., ngVLA). The third row shows the corresponding star formation rate estimate from the total IR luminosity (i.e., $y$-axis label on the right) derived from \texttt{CIGALE}. 

Figure~\ref{fig:compDSFG_lowf} shows the case of low frequencies available in SKA (0.11, 0.3, 0.77, 1.4, 6.7, 12.5 GHz) and in ngVLA (1.4, 6.7, 12.5 GHz). Unlike the UV-bright star-forming galaxies, a massive DSFG with M$_{*}=10^{11}$M$_{\odot}$ is bright in radio--FIR frequency and can be detected in $\nu > 1$ GHz with $>3\sigma$ significance up to $z\sim$5--7 regardless of the dust temperature: dust temperature does not change the radio SED shape except in high-frequency with a notieable difference shown in 12.5GHz. However, since such massive galaxy does not exist in high-redshift, it is more useful to trace the detectable minimum stellar mass. The minimum stellar mass of DSFGs for them to be detected in each observing frequency increases with redshift, as shown in the panels of the second row. Note that, in the panels of the second row, the redshift for the minimum stellar mass of $10^{11}$M$_{\odot}$ is where the flux density variation in the panels of the first row meets the telescope sensitivity limit ($3\sigma$ sensitivity in 1 hour of integration). With increasing observing frequency, the radio telescope is more sensitive to fainter galaxies. For $\nu \gtrsim 1$ GHz observing frequency, M$_{*}=10^{10}$M$_{\odot}$ DSFGs are observable ($>3\sigma$ detection) up to $z\sim$5--7 and M$_{*}=10^{9}$M$_{\odot}$ DSFGs are observable up to $z\sim$2--3. For $\nu \lesssim 1$ GHz observing frequency, M$_{*}=10^{10}$M$_{\odot}$ DSFGs are observable ($>3\sigma$ detection) up to $z\sim 2$.

Figure \ref{fig:compDSFG_highf} shows the case of high frequencies available in ngVLA (27, 41, 93 GHz) and in ALMA (93, 150, 250, 350 GHz). The situation becomes even more promising than the lower frequency case: a massive DSFG (shown in the first row) with any assumed dust temperature (20--90K) is detectable at all redshifts by ngVLA and ALMA. The minimum stellar mass (in the second row) and the corresponding star formation rate (in the third row) for DSFG to be detected in high frequency is more than an order of magnitude lower than that in low frequency. In Figure~\ref{fig:compDSFG_highf}, we add additional plots in the fourth row showing the minimum dust mass of DSFGs for a range of dust temperature (20--90 K) given the IR luminosity that produces a flux density larger than the $3\sigma$ sensitivity limit. We note that the dust mass required for $>3\sigma$ detection is $10^8$M$_{\odot}$ or smaller if the dust temperature, $T_{d,*}\gtrsim30$K and this dust mass is smaller than the corresponding $3\sigma$ detection stellar mass (shown in the panels in the second row) for $T_{d,*}\gtrsim30$K (M$_{*}\sim10^8$M$_{\odot}$). However, for cold dust ($T_{d,*}<30$ K), the required dust mass for $3\sigma$ detection can be as large as $10^{9-10}$M$_{\odot}$, which is comparable to or larger than the corresponding stellar mass for $3\sigma$ detection (in the second row). It implies that for the given IR luminosity, the cold dust may need to have an unrealistically large mass for detection. Also, the CMB emission at $z\sim 10$ already has a higher temperature ($\sim 30$ K) than that of cold dust ($T_{d,*}<30$ K) and hides the thermal dust emission. Therefore the detected DSFGs are unlikely to have cold dust temperature (10--20K). 

\section{Discussion}\label{sec:discussion}
The main result of this work is to predict the galaxy radio-FIR SED for two main galaxy populations: UV-bright star-forming galaxies and DSFGs, and investigate the prospect of observing high-$z$ galaxies in observing frequency from radio to FIR in the era of next-generation radio telescopes. To demonstrate the fidelity of our work, we compare the predictions of the galaxy radio-FIR SEDs and their flux density evolution as a function of redshift against the current state-of-the-art deep radio-millimeter continuum observations. Although the current deep radio-millimeter observations are limited by sky area, observing frequency, and sensitivity and thus cannot be directly compared to our predictions in the full parameter range: redshift, observing frequency, stellar mass, we discuss at our best the current deep radio-millimeter observations and their findings in the context of our forecast of the radio-FIR SED for high-$z$ galaxies, which will demonstrate the usefulness of our work in designing the future radio surveys and observing programs and interpreting the observed results.   

\subsection{Deep radio continuum observation}\label{sec:radio_survey}
There is a (not exhaustive) list of deep radio continuum surveys including VLA-COSMOS 1.4GHz Survey \citep{schinnerer_etal_2007}, VLA-COSMOS 3GHz Large Project \citep{smolcic_etal_2017}, VLA-GOODS-North/South observation at 5GHz \citep{gim_etal_2019}, VLA-Frontier Fields Survey at 3 and 6 GHz \citep{heywood_etal_2021}, VLA-GOODS-North 10GHz survey \citep{jimenez-andrade_etal_2024}, MeerKAT DEEP2 survey at 1.28 GHz \citep{mauch_etal_2020}, MIGHTEE continuum survey at 1.2--1.3 GHz \citep{hale_etal_2025}, EMU survey at 1.4 GHz \citep{hopkins_etal_2025}, LOFAR Two-meter Sky Survey at 150 MHz\citep{tasse_etal_2021}, several GMRT and uGMRTsurveys \citep[e.g.,][]{orcan_etal_2020,sinha_etal_2023}. 

The SFR-M$_{*}$ relation for the galaxies in the VLA-COSMOS 3GHz Large Project is measured up to $z=6$ and the high-end of stellar mass range is $10^{11}$M$_{\odot}$ \citep{leslie_etal_2020}, which is comparable to the range of galaxy properties explored in our analysis of the low frequency observation (Figure~\ref{fig:compJAGUAR_lowf} and \ref{fig:compDSFG_lowf}). The 1.4 GHz radio luminosity distribution from the VLA-COSMOS 3GHz Large Project as a function of redshift has a luminosity limit L$_{\mbox{\tiny 1.4 GHz}}=10^{24.5}$ [W Hz$^{-1}$] for 5$\sigma$ detection with 2.3$\mu$Jy beam$^{-1}$ RMS at $z=5$ \citep{novak_etal_2017}. Similarly, the 1.3 GHz radio luminosity distribution from the MIGHTEE continuum survey as a function of redshift has a luminosity limit L$_{\mbox{\tiny 1.3 GHz}}=10^{24}$ [W Hz$^{-1}$] for 5$\sigma$ detection with 3.6$\mu$Jy beam$^{-1}$ RMS at $z=5$ \citep{pinjarkar_etal_2025}. This radio luminosity limit (L$_{\nu}\approx10^{24}$ [W Hz$^{-1}$] at 1.3--1.4 GHz) corresponds to the observed flux density of 3.8 $\mu$Jy at $z=5$, which is comparable to the expected flux density at 1.4 GHz (1.2$\mu$Jy) for our simulation of the UV-bright star-forming galaxies with M$_{*}=10^{10.5}$M$_{\odot}$ in Figure~\ref{fig:compJAGUAR_lowf} and significantly lower than the expected flux density ($\approx20 \mu$Jy) at 1.4 GHz for our simulation of DSFGs with M$_{*}=10^{11}$M$_{\odot}$ in Figure~\ref{fig:compDSFG_lowf}. The predicted model flux density at 1.4 GHz from our simulated SEDs is larger than the flux density limits of these observations with the similar observing frequencies for detecting galaxies with similar stellar mass range, which suggests that the model prediction is consistent with the observations. 

The VLA-Frontier Fields Survey has 0.9$\mu$Jy beam$^{-1}$ RMS for both 3 and 6 GHz and probes star-forming galaxies with median stellar mass of $10^{10.4}$M$_{\odot}$ at $0.3<z<3$ \citep{jimenez-andrade_etal_2021}. The expected flux density of UV-bright star-forming galaxies in $10^{10.5}$M$_{\odot}$ stellar mass bin at 6.7GHz observing frequency (Figure~\ref{fig:compJAGUAR_lowf}) decrease from $\approx 4\mu$Jy at $z=1$ to $\approx 1\mu$Jy at $z=4$, which is larger than the $1\sigma$ RMS sensitivity of the VLA-Frontier Fields Survey at 6GHz. Although the statistical significance of the expected flux density at 6.7 GHz is weak compared to $1\sigma$ RMS of the 6 GHz VLA-Frontier Fiels Survey (0.9$\mu$Jy beam$^{-1}$ RMS), the expected radio flux density is indeed larger if we use the `version 2' simulation where we consider both synchrotron dimming and redshift-dependent magnetic field enhancement to capture the observed trend of weak evolution of radio--IR correlation as discussed in Section~\ref{subsec:synchphysics}. Like the predicted model flux density at 1.4 GHz, the model flux density at around 6 GHz is also consistent with the observations detecting galaxies with similar stellar mass and redshift range.

In lower frequency observation ($\nu \lesssim 1$ GHz), a clean separation of SF from AGN in radio flux is not easy but several studies characterized the galaxy population seen in the low frequency deep radio images based on the best effort of SED modeling using cross-matched multiwavelength photometry.  

The GMRT survey at 610 MHz detects star-forming galaxies up to $z\simeq1.8$ from ELAIS N1 field with $\approx 7.1\mu$Jy beam$^{-1}$ RMS \citep{orcan_etal_2020} and the uGMRT survey at 400 MHz detects star-forming galaxies up to $z\approx5$ from Bo\"otes field with $\approx 35\mu$Jy beam$^{-1}$ RMS \citep{sinha_etal_2023}. For UV star-forming galaxies with M$_{*}=10^{10.5}$M$_{\odot}$ in Figure~\ref{fig:compJAGUAR_lowf}, the predicted flux densities at observing frequency of 770 MHz (close to 610) and 300 MHz (close to 400 MHz) are larger than the sensitivities of these observations at the maximum redshift that they can achieve ($z=1.8$ for 610 MHz observation and $z=5$ for 400 MHz observation).   

The LOFAR observation at 144 MHz detects bright sub-\textit{mm} galaxies (M$_{*}=10^{11-12}$M$_{\odot}$) at $1.7<z<3.5$ with a range of the integrated flux density: 100--1000 $\mu$Jy \citep{bondi_etal_2025}. Our predicted flux density of DSFGs at 110 MHz for M$_{*}=10^{11}$M$_{\odot}$ bin is $>100$ $\mu$Jy up to $z=5$ as seen in Figure~\ref{fig:compDSFG_lowf}, which is consistent with the low frequency observation of sub-\textit{mm} galaxies \citep{bondi_etal_2025}. Also the 150MHz observation of the LOFAR deep fields \citep{cochrane_etal_2023} with 20 $\mu$Jy beam$^{-1}$ RMS reveals star-forming galaxies up to $z=4$ and the recent re-analysis of the data in \cite{cochrane_etal_2023} detects galaxies up to $z=5.7$ above 100-110 $\mu$Jy ($5\sigma$ significance) flux density \citep{wang_etal_2025}. These high-$z$ galaxies ($z=4-5$) have high SFR ($>100$M$_{\odot}$ yr$^{-1}$) and radio luminosity (L$_{\mbox{\tiny 150 MHz}}>10^{24}$ [W Hz$^{-1}$]) \citep{cochrane_etal_2023} and are likely to be high mass (M$_{*}>10^{10}$M$_{\odot}$) star-forming galaxies based on the stellar mass dependent correlation between radio luminosity and SFR derived from LOFAR Deep Fields data \citep{smith_etal_2021}. Our predicted flux density at 110 MHz for UV-bright star-forming galaxies (Figure~\ref{fig:compJAGUAR_lowf}) and DSFGs (Figure~\ref{fig:compDSFG_lowf}) with high stellar mass (M$_{*}>10^{10-11}$M$_{\odot}$) is comparable to or larger than the sensitivity limit of the LOFAR deep survey. 

\subsection{ALMA continuum observation}\label{sec:mm_survey}
Most deep millimeter/FIR continuum observations of high-$z$ galaxies are performed by aggregating continuum spectrum from large spectroscopic surveys by ALMA \citep{gonzalez-lopez_etal_2020,mitsuhashi_etal_2024b,inami_etal_2022} while early ALMA operation imaged Hubble Ultra Deep Field (HUDF) with continuum spectral setting \citep{mclure_etal_2018}.

A stacking analysis of the star-forming galaxies at $2<z<3$ in the ALMA 1.3\textit{mm} deep continuum observation of HUDF  \citep{mclure_etal_2018} does not detect galaxies with M$_{*} = 10^{8.5-9.25}$M$_{\odot}$ for $1\sigma$ RMS of 3$\mu$Jy achieved by stacking. However, more massive galaxies with M$_{*}=10^{9.25-10}$M$_{\odot}$ are detected by stacking. The expected flux density in the 250 GHz (similar to 1.3\textit{mm}) ALMA band for UV-bright star-forming galaxies in Figure~\ref{fig:compJAGUAR_highf} for $\mbox{M}_{*}=10^{9.5}$M$_{\odot}$ mass bin ($\gtrsim 20 \mu$Jy for a reasonable dust temperature, 30K) is larger than  5$\sigma$ detection based on 3$\mu$Jy beam$^{-1}$ RMS from stacking and explains the detection of the stacked M$_{*}=10^{9.25-10}$M$_{\odot}$ galaxies in \cite{mclure_etal_2018}.

Our predictions of the expected flux density in the 250 GHz ALMA band for UV-bright star-forming galaxies with $\mbox{M}_{*}=10^{9.5-10.5}$M$_{\odot}$ in Figure~\ref{fig:compJAGUAR_highf} are also consistent with the ALMA detection of dust continuum emission at a level of 50--100 $\mu$Jy in Band 7 and 9 from UV-bright star-forming galaxies with the similar stellar masses at $4<z<6$ in the ALMA-CRISTAL Survey \citep{mitsuhashi_etal_2024b} and in Band 3 (230-300 GHz) from UV-bright star-forming galaxies with the similar stellar masses at $6<z<8$ in the REBELS survey \citep{inami_etal_2022}. Our prediction also explains the FIR non-detection of the similar star-forming galaxy population with lower stellar mass ($\mbox{M}_{*}\sim10^{9}$M$_{\odot}$) in higher redshift ($z\gtrsim10$) for individual system \citep[e.g.,][]{bakx_etal_2023,fujimoto_etal_2023,popping_2023,yoon_etal_2023} and even for stacked deep image of all available ALMA observations of the galaxies at $z>8.3$ observed so far \citep{bakx_etal_2026}.   

\section{Summary} \label{sec:summary}
This study presents the prospects of observing radio--FIR continuum emission from galaxies for a wide range of redshifts, using the current (ALMA) and the next-generation (SKA and ngVLA) high-sensitivity radio telescopes. Based on the assumed stellar population parameters, we create panchromatic galaxy SEDs from UV/optical to FIR--radio by including the CMB effect and adopting the radio--IR correlation. Two populations of galaxies were modeled --- (1) the JAGUAR mock catalog simulating the JWST-discovered, UV-bright star-forming galaxies and (2) dusty star-forming galaxies. Although we compare the SED to the nominal sensitivities of SKA, ngVLA and ALMA at the several chosen frequencies, our investigation is generally applicable to any observing frequencies in radio-FIR. We summarize our findings as follows.

\begin{enumerate}
\item[$\bullet$] The SED of Anomalous Microwave Emission (AME) is a minor and almost negligible contribution to the radio SED based on spatially integrated flux measurement if $z>0.1$ (see Figure~\ref{fig:pflux}).

\item[$\bullet$] CMB changes the SED shape in FIR wavelength by dust heating and background contrast (see Figure~\ref{fig:zcmb}) such that the shift of the FIR SED peak frequency due to the variation of the assumed dust temperature, which is significant in the local Universe, becomes small in the high-redshift Universe (see Figure~\ref{fig:temp}).  

\item[$\bullet$] For UV-bright star-forming galaxies sampled from the JAGUAR catalog, unless the galaxies are extremely massive (M$_*\gtrsim 10^{11}$M$_{\odot}$), they are not likely to be detected in radio ($\nu \lesssim 10$GHz) as seen in Figure~\ref{fig:compJAGUAR_lowf}. Apparently, the most promising observing frequency to detect the UV-bright high-$z$ star-forming galaxies with M$_*\gtrsim 10^{9.5}$M$_{\odot}$ is $\approx 100$ GHz which has a benefit from negative $K$-correction and is still below the rest-frame FIR peak of thermal dust emission (see Figure~\ref{fig:compJAGUAR_highf}).

\item[$\bullet$] Dusty star-forming galaxies (DSFGs) are bright in radio--FIR. For low frequency range (available in the SKA bands and the ngVLA low-frequency bands), DSFGs with M$_*\gtrsim 10^{10}$M$_{\odot}$ can be detected up to $z\sim 2$ in $\nu \lesssim 1 $GHz and up to $z \sim$5--7 in $\nu \gtrsim 1 $GHz (Figure~\ref{fig:compDSFG_lowf}). For high frequency range (available in the ngVLA high-frequency bands and the ALMA bands), M$_*\sim 10^{9}$M$_{\odot}$ can be easily detected up to $z\sim10$ and even smaller galaxies can be detected regardless of the redshift in 93 GHz (Figure~\ref{fig:compDSFG_highf}).

\item[$\bullet$] Overall, the most promising workhorse observing window for high-$z$ galaxy observation is $\approx 100$ GHz observing frequency: in principle, it is capable of detecting galaxies with M$_*\gtrsim 10^{9}$M$_{\odot}$ for any redshift up to the epoch of first galaxy formation, whether or not they are UV-bright star forming galaxies or DSFGs.

\end{enumerate}
In the era of the JWST discovery of large populations of high-$z$ galaxies ($z>10$), current state-of-the-art and next-generation radio telescopes with the capability of detecting such high-$z$ galaxies in radio--FIR will provide excellent complementary observation. Ultimately, due to the redshift-independent flux density, we may be able to detect the first galaxies just as they emerge from the cosmic dark age, using the next-generation radio telescopes. 

\begin{acknowledgments}
We thank our anonymous referee who provided constructive feedback that improved the paper. This work is based on a program selected for the ngVLA Community Studies Program (Round 5). IY thanks Chris Carilli and Roberto Decarli for thoughtful comments on the early stage of the work. E.F.-J.A. acknowledge support from UNAM- PAPIIT project IA104725, and from CONAHCyT Ciencia de Frontera project ID: CF-2023-I- 506. The National Radio Astronomy Observatory is a facility of the National Science Foundation operated under cooperative agreement by Associated Universities, Inc.
\end{acknowledgments}

\vspace{5mm}
\facilities{}

\software{CIGALE \citep{Boquien_2019}, NumPy \citep{numpy_2020}, Matplotlib \citep{matplotlib_2007}}

\appendix
\section{Galaxies in the JAGUAR mock catalog}
\begin{figure}[h]
    \centering
    \gridline{
    \fig{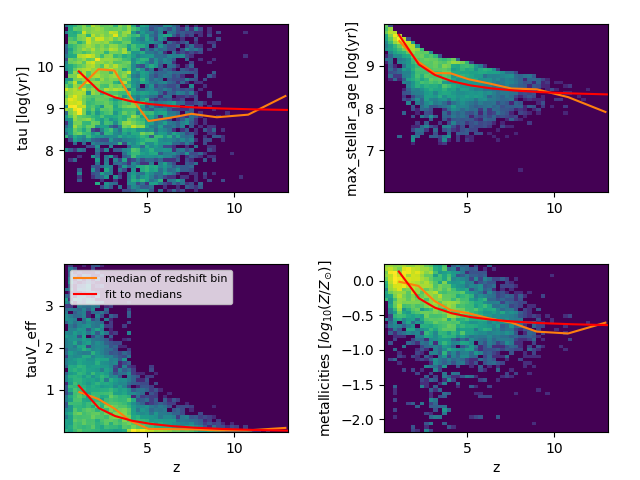}{3.2in}{$M_{char}(z)$}
    \fig{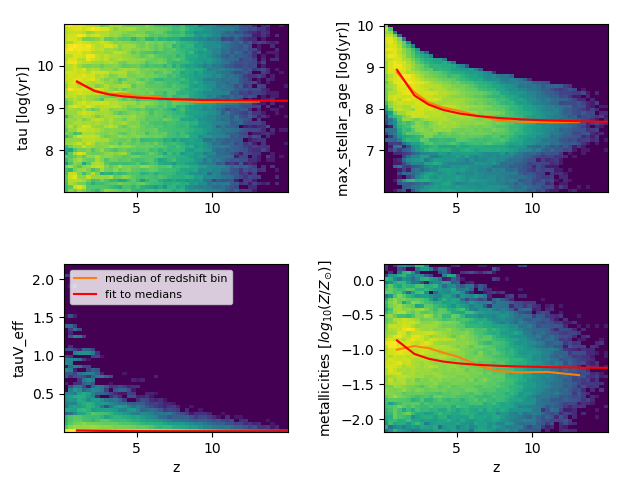}{3.2in}{$6\leq\mathcal{M}<7$}
    }
    \gridline{
    \fig{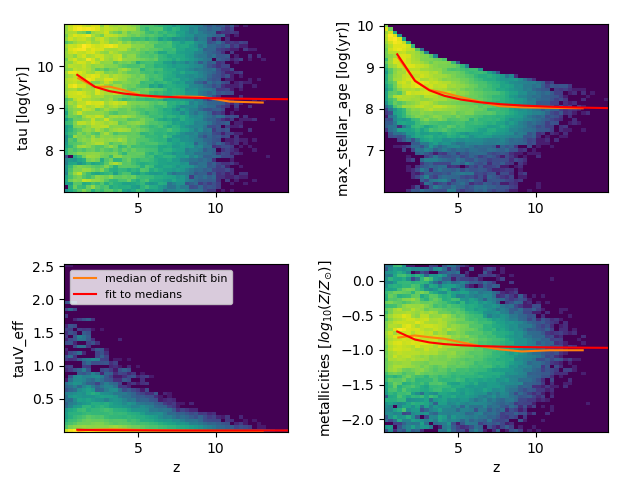}{3.2in}{$7\leq\mathcal{M}<8$}
    \fig{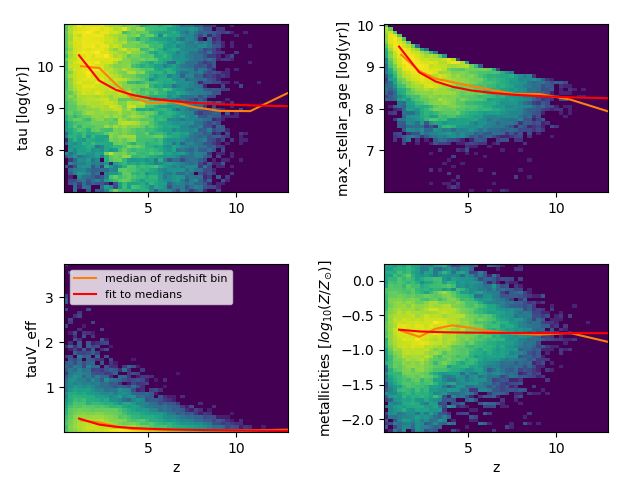}{3.2in}{$8\leq\mathcal{M}<9$}
    }
    \gridline{
    \fig{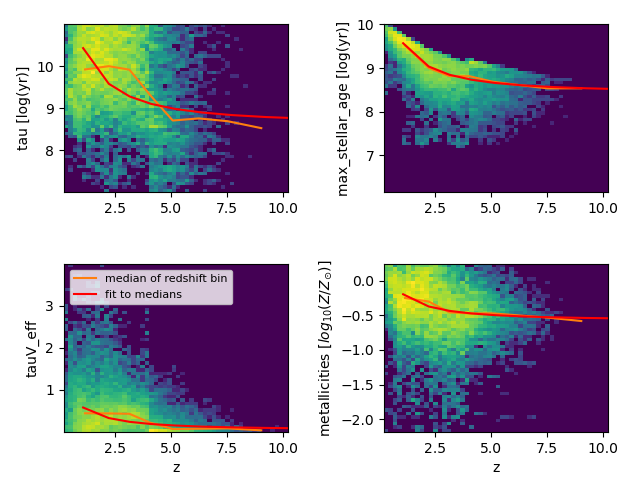}{3.2in}{$9\leq\mathcal{M}<10$}
    \fig{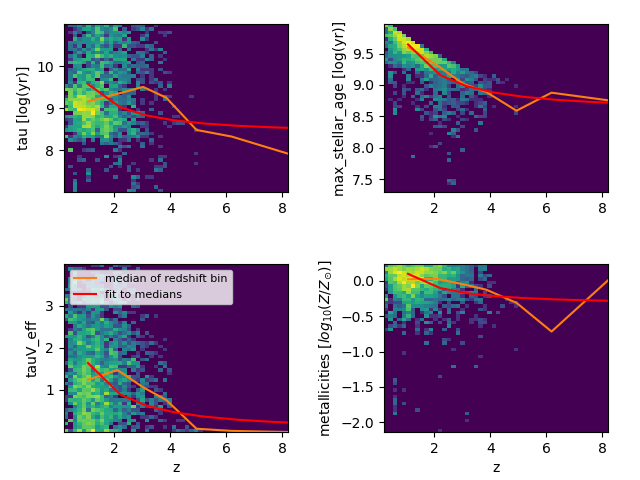}{3.2in}{$10\leq\mathcal{M}<11$}
    }
    \caption{Each group of four panels shows a 2-D histogram overlaid with the fits to the four physical parameters of the galaxies in each mass bin of the JAGUAR catalog. The orange lines indicate the median values at each redshift bin, while the red lines show the fit to these median values as a function of time. Note that the x-axis is re-scaled to redshift. Also note the changes in axes' limits and scale.}
    \label{fig:JAGUARfits}
\end{figure}

For redshift and stellar mass bins, we sample galaxies from the JAGUAR catalog and use their average stellar population parameters, as explained in Section~\ref{subsec:JAGUAR}. Here we provide detailed information regarding the number of galaxies in each stellar mass and redshift bin in Table~\ref{tab:counts} and the result of fitting the distribution of stellar population parameters to extrapolate to the higher redshift, shown in Figure~\ref{fig:JAGUARfits}. 

%\begin{rotatetable}
\begin{splitdeluxetable}{|l|c|c|c|c|B|c|c|c|c|c|c|}
%\begin{splitdeluxetable}{lccccBcccccc}
\label{tab:counts}
\tabletypesize{\footnotesize}
\tablewidth{0pt}
\tablecaption{Number counts for each redshift and mass bin of the JAGUAR mock catalog}
\startdata
$[\mathcal{M}_{l},\mathcal{M}_{u})$\textbackslash$(z_{l}$,$z_{u}]:z_{mean}$ & (0.0,1.760]:1.057 & (1.760,2.689]:2.207 & (2.689,3.628]:3.140 & (3.628,4.647]:4.109 & (4.647,5.789]:5.177 & (5.789,7.094]:6.384 & (7.094,8.606]:7.739 & (8.606,10.378]:9.263 & (10.378,12.480]:11.093 & (12.480,15.0]:13.221\\\hline
[6,7) & 718455 & 375592 & 289299 & 239316 & 197319 & 149282 & 95882 & 32756 & 6756 & 865\\\hline
[7,8) & 197283 & 131198 & 107669 & 87805 & 60721 & 37565 & 19330 & 5188 & 764 & 67\\\hline
[8,9) & 56930 & 44888 & 38595 & 30230 & 17189 & 8435 & 3348 & 614 & 67 & 2\\\hline
[9,10) & 17247 & 12307 & 9666 & 6612 & 2662 & 870 & 176 & 18 & 0 & 0\\\hline
[10,11) & 5540 & 1738 & 508 & 126 & 27 & 1 & 1 & 0 & 0 & 0\\\hline
[11,$\infty$) & 279 & 112 & 15 & 4 & 1 & 0 & 0 & 1 & 0 & 0\\\hline
$\mathcal{M}_{char}$ & 8584 & 5327 & 4882 & 4260 & 2507 & 1415 & 667 & 185 & 38 & 3\\\tableline
\enddata
\tablecomments{Redshift bins beyond $z=15$ are not shown because there are no galaxies in the JAGUAR catalog beyond $z=15$. Those galaxies that were the only ones in their redshift/mass bin (i.e. wherever there is a ``1'') often had physical properties that were quite different from those in ``nearby'' bins.}
\end{splitdeluxetable}
%\end{rotatetable}

%\begin{splitdeluxetable*}{l|c|c|c|c|Bc|c|c|c|c|c|}
%\label{tab:counts}
%\tabletypesize{\scriptsize}
%\tablewidth{0pt}
%\tablecaption{Number counts for each redshift and mass bin of the JAGUAR mock catalog }
%\startdata
%\colhead{$[\mathcal{M}_{lower},\mathcal{M}_{upper})$\textbackslash$(z_{lower}$,$z_{upper}]:z_{mean}$} & (0.0,1.760]:1.057 & (1.760,2.689]:2.207 & (2.689,3.628]:3.140 & (3.628,4.647]:4.109 & (4.647,5.789]:5.177 & (5.789,7.094]:6.384 & (7.094,8.606]:7.739 & (8.606,10.378]:9.263 & (10.378,12.480]:11.093 & (12.480,15.0]:13.221\\\hline
%[6,7) & 718455 & 375592 & 289299 & 239316 & 197319 & 149282 & 95882 & 32756 & 6756 & 865\\\hline
%[7,8) & 197283 & 131198 & 107669 & 87805 & 60721 & 37565 & 19330 & 5188 & 764 & 67\\\hline
%[8,9) & 56930 & 44888 & 38595 & 30230 & 17189 & 8435 & 3348 & 614 & 67 & 2\\\hline
%[9,10) & 17247 & 12307 & 9666 & 6612 & 2662 & 870 & 176 & 18 & 0 & 0\\\hline
%[10,11) & 5540 & 1738 & 508 & 126 & 27 & 1 & 1 & 0 & 0 & 0\\\hline
%[11,$\infty$) & 279 & 112 & 15 & 4 & 1 & 0 & 0 & 1 & 0 & 0\\\hline
%$\mathcal{M}_{char}$ & 8584 & 5327 & 4882 & 4260 & 2507 & 1415 & 667 & 185 & 38 & 3\\\hline
%\enddata
%\tablecomments{Redshift bins beyond $z=15$ are not shown because there are no galaxies in the JAGUAR catalog beyond $z=15$. Those galaxies that were the only ones in their redshift/mass bin (i.e. wherever there is a ``1'') often had physical properties that were quite different from those in ``nearby'' bins.}
%\end{splitdeluxetable*}

\bibliography{main}{}
\bibliographystyle{aasjournal}

\end{document}